\documentclass[12pt]{article}
\usepackage{graphicx}
\usepackage{epstopdf}
\usepackage{amsmath,amssymb,amsfonts,graphicx}

\newcommand{\bnabla}{\mbox{\boldmath $\nabla$}}

\begin{document}

\thispagestyle{empty}
\renewcommand{\refname}{References}

\title{\bf Pressure from the vacuum of confined 
spinor matter}

%\maketitle

\author{Yu. A. Sitenko$^{1,2}$ and S. A. Yushchenko$^{1}$}

\date{}

\maketitle
\begin{center}
$^{1}$ Bogolyubov Institute for Theoretical Physics,\\
National Academy of Sciences of Ukraine,\\
14-b Metrologichna Str., 03680 Kyiv, Ukraine\\
$^{2}$ Institute for Theoretical Physics, University of Bern,\\
Sidlerstrasse 5, CH-3012 Bern, Switzerland
\end{center}

%\ead{yusitenko@bitp.kiev.ua}

\begin{abstract}
Charged spinor matter field is quantized in a spatial
region bounded by two parallel neutral plates. The most general set of
boundary conditions ensuring the confinement of matter within the plates 
is considered. We study a response of the vacuum of the confined matter to the 
background uniform magnetic field which is directed orthogonally to the plates. 
It is proven that, in the case of a sufficiently strong magnetic
field, the vacuum pressure onto the plates is positive and independent of the boundary
condition, as well as of the distance between the plates.
\end{abstract}


\bigskip

\begin{center}
\end{center}

\bigskip
\medskip

\section{Introduction}

Perhaps, a quest for boundary conditions ensuring the confinement of the quantized spinor matter was initiated in 
the context of a model description of hadrons as composite systems with their internal structure being associated 
with quark-gluon constituents \cite{Cho1,Cho2,De,Joh}. If an hadron is an extended object occupying spatial region $\Omega$
bounded by surface $\partial{\Omega}$, then the condition that the quark matter field be confined inside the hadron 
is formulated as
$$
\boldsymbol{n}\cdot\boldsymbol{J}(\mathbf{r})|_{\mathbf{r}
\in
\partial{\Omega}}=0,\eqno(1)
$$
where $\boldsymbol{n}$ is the unit normal to the boundary surface, and 
$\boldsymbol{J}(\mathbf{r})={\psi}^{\dag}(\mathbf{r})\boldsymbol{\alpha}\psi(\mathbf{r})$ with $\psi(\mathbf{r})$ ($\mathbf{r}
\in \Omega$) being the quark matter field (${\alpha}^1$, ${\alpha}^2$, ${\alpha}^3$ and ${\beta}$ are the generating elements of 
the Dirac-Clifford algebra); an appropriate condition is also formulated for the gluon matter field.

The concept of confined matter fields is quite familiar in the context of condensed matter physics: collective excitations 
(e.g., spin waves and phonons) exist only inside material objects and do not spread outside. Moreover, in the context 
of quantum electrodynamics, if one is interested in the effect of a classical background magnetic field on the vacuum of 
the quantized electron-positron matter, then the latter should be considered as confined to the spatial region between 
the sources of the magnetic field, as long as collective quasielectronic excitations inside a magnetized material differ from 
electronic excitations in the vacuum. It should be noted in this respect that the study of the effect of the
background electromagnetic field on the vacuum of quantized charged matter has begun already eight decades ago
\cite{Hei1,Eul,Hei2,Wei,Schw}, see review in \cite{Dun}. However, the concern has been for the case of a background field 
filling the whole (infinite) space, that is hard to be regarded as realistic. The case of both the background and 
quantized fields confined to a bounded spatial region with boundaries serving as sources of the background field looks 
much more physically plausible, it can even be regarded as realizable in laboratory. Moreover, there is no way to detect 
the energy density that is induced in the vacuum in the first case, whereas the pressure from the vacuum onto the boundaries, 
resulting in the second case, is in principle detectable.

In view of the above, an issue of a choice of boundary conditions for the quantized matter fields gains a crucial
significance, and condition (1) should be resolved to take the form of a boundary condition that is linear in $\psi(\mathbf{r})$. Recall 
that an immediate way of such a resolution is known as the MIT bag boundary condition \cite{Joh},
$$
[I + {\rm i}\beta(\boldsymbol{n}\cdot\boldsymbol{\alpha})]\psi(\mathbf{r})|_{\mathbf{r}\in
\partial{\Omega}}=0, \eqno(2)
$$
but it is needless to say that this way is not a unique one. The most general boundary condition that is linear in 
$\psi(\mathbf{r})$ in the case of a simply-connected boundary involves four arbitrary parameters \cite{Wie}, and the explicit form of this boundary condition has been given \cite{Si1}. The 
condition is compatible with the self-adjointness of the differential operator of one-particle energy in first-quantized 
theory (Dirac hamiltonian operator in the case of relativistic spinor matter). The self-adjointness of operators of physical observables 
is required by general principles of comprehensibility and mathematical consistency, see, e.g., \cite{Neu,Akhi}. To put it 
simply, a multiple action is well defined for a self-adjoint operator only, allowing for the construction of functions of 
the operator, such as resolvent, evolution, heat kernel and zeta-function operators, with further implications upon second
quantization. In the present paper, we follow the lines of works \cite{Wie, Si1} by proposing a different, 
embracing more cases, form of the four-parameter generalization of the MIT bag boundary condition.

Thus, we consider in general the quantized spinor matter field that is confined to the
three-dimensional spatial region $\Omega$ bounded by the two-dimensional surface $\partial{\Omega}$. To study 
a response of the vacuum to the background magnetic field, we restrict ourselves to the case of the boundary 
consisting of two parallel planes; the magnetic field is assumed to be uniform and orthogonal to the planes. 
Such a spatial geometry is typical for the remarkable macroscopic quantum phenomenon which yields the attraction 
(negative pressure) between two neutral plates and which is known 
as the Casimir effect \cite{Cas1}, see reviews in \cite{Mil,Bor}. The conventional Casimir effect is due to vacuum 
fluctuations of the quantized electromagnetic field obeying certain boundary conditions at the bounding plates, and 
a choice of boundary conditions is physically motivated by material properties of the plates (for instance, metallic 
or dielectric ones, see, e.g., \cite{Bor}). Such a motivation is lacking for the case of vacuum fluctuations of the 
quantized spinor matter field. That is why there is a necessity in the last case to start from the most general set of 
mathematically acceptable (i.e. compatible with the self-adjointness) boundary conditions. Then follows, as has been already 
discussed, a physical constraint that spinor matter be confined within the plates. A further physical constraint, as will be shown 
in Section 4, is that the spectrum of the wave number vector in the direction which is orthogonal to the plates be real and 
unambiguously (although implicitly) determined. Employing these mathematical and physical restrictions, we explore the 
generalized Casimir effect which is due to vacuum fluctuations of the quantized spinor matter field in the presence of the 
background magnetic field; the pressure 
from the vacuum onto the bounding plates will be found.

In the next section we show how the requirement of the self-adjointness for the Dirac hamiltonian operator brings 
the most general set of boundary conditions ensuring the confinement of the quantized spinor matter in the cases of 
a simply-connected boundary and a disconnected boundary consisting of two noncompact noncontiguous surfaces. In Section 3
we consider the vacuum energy which is induced by a background uniform magnetic field in the cases of the unbounded 
quantization volume and the quantization volume bounded by two parallel infinite plates. The boundary condition
determining unambiguously the spectrum of the wave number vector in the direction orthogonal to the plates is derived, and 
the general expression for the pressure from the vacuum onto the plates is obtained in Section 4. The vacuum pressure in 
some particular cases is examined in Section 5, while the asymptotical behaviour of the vacuum pressure at small and large 
separations of the plates is analysed in Section 6. Finally, the results are summarized and discussed in Section 7. We 
adduce the solution to the Dirac equation in the background uniform magnetic field in Appendix A. In Appendix B we derive 
a new version of the Abel-Plana formula for summation over values of the wave number vector in the direction orthogonal to 
the plates.

\section{Self-adjointness and boundary conditions}

Defining a scalar product as $(\tilde{\chi},\chi)=\int\limits_{\Omega}{\rm
d}^3r\,\tilde{\chi}^{\dag}\chi$, we get, using integration by parts,
$$
(\tilde{\chi},H\chi)=(H^{\dag}\tilde{\chi},\chi)- {\rm
i}\int\limits_{\partial{\Omega}}{\rm
d}\mathbf{s}\cdot\tilde{\chi}^{\dag}\boldsymbol{\alpha}\chi,\eqno(3)
$$
where
$$
H=H^{\dag}=-{\rm
i}\boldsymbol{\alpha}\cdot\bnabla+\beta{m} \eqno(4)
$$
is the formal expression for the Dirac hamiltonian operator and ${\bnabla}$ is the covariant derivative involving both
the affine and bundle connections (natural units ${\hbar}=c=1$ are used). Operator $H$ is
Hermitian (or symmetric in mathematical parlance),
$$
(\tilde{\chi},H\chi)=(H^{\dag}\tilde{\chi},\chi),\eqno(5)
$$
if
$$
\int\limits_{\partial{\Omega}}{\rm
d}\mathbf{s}\cdot\tilde{\chi}^{\dag}\boldsymbol{\alpha}\chi
= 0. \eqno(6)
$$
The latter condition can be satisfied in various ways by imposing
different boundary conditions for $\chi$ and $\tilde{\chi}$.
However, among the whole variety, there may exist a possibility
that a boundary condition for $\tilde{\chi}$ is the same as that
for $\chi$; then the domain of definition of $H^{\dag}$ (set of
functions $\tilde{\chi}$) coincides with that of $H$ (set of
functions $\chi$), and operator $H$ is self-adjoint. The
action of a self-adjoint operator results in functions belonging
to its domain of definition only, and, therefore, a multiple
action and functions of such an operator can be consistently defined.

Condition (6) is certainly fulfilled when the integrand in (6)
vanishes, i.e.
$$
\tilde{\chi}^{\dag}(\boldsymbol{n}\cdot\boldsymbol{\alpha})\chi|_{\mathbf{r}
\in \partial{\Omega}}=0. \eqno(7)
$$
To fulfill the latter condition, we impose the same
boundary condition for $\chi$ and $\tilde{\chi}$ in the form
$$
\chi|_{\mathbf{r}\in \partial{\Omega}}=K\chi|_{\mathbf{r}\in
\partial{\Omega}},\quad \tilde{\chi}|_{\mathbf{r}\in
\partial{\Omega}}=K\tilde{\chi}|_{\mathbf{r}\in \partial{\Omega}},\eqno(8)
$$
where $K$ is a matrix (element of the Dirac-Clifford algebra) which is
determined by two conditions:
$$
K^{2}=I \eqno(9)
$$
and
$$
K^{\dag}(\boldsymbol{n}\cdot\boldsymbol{\alpha})K=-\boldsymbol{n}\cdot\boldsymbol{\alpha}. \eqno(10)
$$
It should be noted that, in addition to (7), the following
combination of $\chi$ and $\tilde{\chi}$ is also vanishing at the
boundary:
$$
\tilde{\chi}^{\dag}(\boldsymbol{n}\cdot\boldsymbol{\alpha})K\chi|_{\mathbf{r}
\in
\partial{\Omega}}=\tilde{\chi}^{\dag}K^{\dag}(\boldsymbol{n}\cdot\boldsymbol{\alpha})\chi|_{\mathbf{r}
\in \partial{\Omega}}=0. \eqno(11)
$$
Using the standard representation for the Dirac matrices,
$$
\beta=\begin{pmatrix}
I&0\\
0&-I
\end{pmatrix},\qquad
\boldsymbol{\alpha}=\begin{pmatrix}
0&\boldsymbol{\sigma}\\
\boldsymbol{\sigma}&0
\end{pmatrix}
$$
($\sigma^{1},\sigma^{2}$ and $\sigma^{3}$ are the Pauli matrices),
one can get
$$
K=\begin{pmatrix}
0&{\varrho}^{-1}\\
\varrho&0
\end{pmatrix},\eqno(12)
$$
where condition
$$
(\boldsymbol{n}\cdot\boldsymbol{\sigma})\varrho=-{\varrho}^{\dag}(\boldsymbol{n}\cdot\boldsymbol{\sigma})
\eqno(13)
$$
defines $\varrho$ as a rank-2 matrix depending on four arbitrary
parameters  \cite{Wie}. An explicit form for matrix $K$ is \cite{Si1}
$$
K=\frac{(1+u^2-v^2-{\boldsymbol{t}}^2)\beta+(1-u^2+v^2+{\boldsymbol{t}}^2)I}{2{\rm
i}(u^2-v^2-{\boldsymbol{t}}^2)}(u\boldsymbol{n}\cdot\boldsymbol{\alpha}+v\beta\gamma^{5}-{\rm
i}\boldsymbol{t}\cdot\boldsymbol{\alpha}),\eqno(14)
$$
where $\gamma^{5}={\rm i}\alpha^1\alpha^2\alpha^3$, and
$\boldsymbol{t}=(t^1,t^2)$ is a two-dimensional vector which is
tangential to the boundary, $\boldsymbol{t}\cdot\boldsymbol{n}=0$. Matrix $K$ is Hermitian, $K^{\dag}=K$,
if either
$$
u=1,\quad v=0, \quad \boldsymbol{t}=0, \eqno(15)
$$
or
$$
u=0,\quad v^2+{\boldsymbol{t}}^2=1. \eqno(16)
$$

Using parametrization
$$
u=\cosh\tilde{\vartheta}\cosh\vartheta, \quad
v=\cosh\tilde{\vartheta}\sinh\vartheta\cos\theta,
$$
$$
t^1=\cosh\tilde{\vartheta}\sinh\vartheta\sin\theta\cos\phi, \quad
t^2=\cosh\tilde{\vartheta}\sinh\vartheta\sin\theta\sin\phi,
$$
$$
-\infty<\vartheta<\infty, \quad 0\leq\tilde{\vartheta}<\infty,
\quad 0\leq\theta<\pi, \quad 0\leq\phi<2\pi \eqno(17)
$$
in the case of $u^2-v^2-{\boldsymbol{t}}^2\geq 1$, one gets
$$
K=\frac{\beta(1+\cosh^2\tilde{\vartheta})-I\sinh^2\tilde{\vartheta}}{2{\rm
i}\cosh\tilde{\vartheta}}
$$
$$
\times
 [\boldsymbol{n}\cdot\boldsymbol{\alpha}\cosh\vartheta+\beta\gamma^{5}\sinh\vartheta\cos\theta-{\rm
i}(\alpha^{1}\cos\phi+\alpha^{2}\sin\phi)\sinh\vartheta\sin\theta)],\eqno(18)
$$
while, using parametrization
$$
u=\cosh\tilde{\varsigma}\sinh\varsigma, \quad
v=\cosh\tilde{\varsigma}\cosh\varsigma\cos\theta,
$$
$$
t^1=\cosh\tilde{\varsigma}\cosh\varsigma\sin\theta\cos\phi, \quad
t^2=\cosh\tilde{\varsigma}\cosh\varsigma\sin\theta\sin\phi,
$$
$$
-\infty<\varsigma<\infty, \quad 0\leq\tilde{\varsigma}<\infty,
\quad 0\leq\theta<\pi, \quad 0\leq\phi<2\pi
\eqno(19)
$$
in the case of $u^2-v^2-{\boldsymbol{t}}^2\leq -1$, one gets
$$
K=\frac{\beta\sinh^2\tilde{\varsigma}-I(1+\cosh^2\tilde{\varsigma})}{2{\rm
i}\cosh\tilde{\varsigma}}
$$
$$
\times[\boldsymbol{n}\cdot\boldsymbol{\alpha}\sinh\varsigma+\beta\gamma^{5}\cosh\varsigma\cos\theta-{\rm
i}(\alpha^{1}\cos\phi+\alpha^{2}\sin\phi)\cosh\varsigma\sin\theta];\eqno(20)
$$
here
$$
[\boldsymbol{n}\cdot\boldsymbol{\alpha},\,\alpha^1]_{+}=
[\boldsymbol{n}\cdot\boldsymbol{\alpha},\,\alpha^2]_{+}=[\alpha^1,\alpha^2]_{+}=0.
\eqno(21)
$$
The intermediate case of $-1 \leq u^2-v^2-{\boldsymbol{t}}^2 \leq
1$ is obtained by going over to imaginary values of parameters
$\tilde{\vartheta}$ and  $\tilde{\varsigma}$:
$$
{\rm Re}\tilde{\vartheta}=0, \quad \quad 0\leq {\rm
Im}\tilde{\vartheta}< \pi/2 \quad (0 < u^2-v^2-{\boldsymbol{t}}^2
\leq 1) \eqno(22)
$$
and
$$
{\rm Re}\tilde{\varsigma}=0, \quad \quad 0\leq {\rm
Im}\tilde{\varsigma}< \pi/2 \quad (-1 \leq
u^2-v^2-{\boldsymbol{t}}^2 < 0). \eqno(23)
$$

Parameters $\vartheta, \tilde{\vartheta}$ (or $\varsigma, \tilde{\varsigma}), \theta$ and $\phi$ 
can be interpreted as the self-adjoint extension parameters. It should be emphasized that the values 
of these parameters vary in general from point to point of the
boundary. In this respect the ``number'' of self-adjoint extension
parameters is in fact infinite, moreover, it is not countable but
is of power of a continuum. This distinguishes the case of an
extended boundary from the case of an excluded point (contact
interaction), when the number of self-adjoint extension parameters
is finite, being equal to $n^2$ for the deficiency index equal to
\{$n,n$\} (see, e.g., \cite{Akhi}).

At the points where matrix $K$ is Hermitian, it takes forms
$$
K_{+}=-{\rm i}\beta(\boldsymbol{n}\cdot\boldsymbol{\alpha}) \quad
(u^2-v^2 -{\boldsymbol{t}}^2 = 1) \eqno(24)
$$
and
$$
K_{-}={\rm i}\beta\gamma^{5}\cos\theta+(\alpha^{1}\cos\phi+\alpha^{2}\sin\phi)\sin\theta 
\quad (u^2-v^2 -{\boldsymbol{t}}^2 = -1). \eqno(25)
$$
A transition from $K_{+}$ to $K_{-}$ in the parametric space is
performed with the use of (18) by varying $\tilde{\vartheta}$ from $0$ to
${\rm
i}\pi/2$ and then with the use of (20) by varying $\tilde{\varsigma}$ from ${\rm
i}\pi/2$ to $0$. Matrix $K_{+}$
(24) corresponds to the choice of the standard MIT bag boundary
condition \cite{Joh}, cf (2),
$$
(I-K_{+})\chi|_{\mathbf{r}\in
\partial{\Omega}}=(I-K_{+})\tilde{\chi}|_{\mathbf{r}\in
\partial{\Omega}}=0, \eqno(26)
$$
when relation (11) takes form
$$
\tilde{\chi}^{\dag}\beta\chi|_{\mathbf{r}\in
\partial{\Omega}}=0.\eqno(27)
$$
To elucidate the meaning of the choice that is corresponded with
matrix $K_{-}$ (25), one has to perform a transition from $K_{+}$
to $K_{-}$ in a parametric space with the same set of parameters,
i.e. two different pairs of parameters, ($\vartheta$,
$\tilde{\vartheta}$) and ($\varsigma$, $\tilde{\varsigma}$),
should be changed to a single one, say, ($\varphi$,
$\tilde{\varphi}$). The most natural way is to revoke the condition that $K$ be
off-diagonal, imposing instead the condition that $K$ be Hermitian
(as $K_{+}$ and $K_{-}$ are). Then, in view of relation (9), $K$
is unitary as well, $K^{\dag}=K^{-1}$, and relation (10) is
rewritten as
$$
[K,\boldsymbol{n}\cdot\boldsymbol{\alpha}]_{+}=0.\eqno(28)
$$
One can simply go through 16 linearly independent elements of the
Dirac-Clifford algebra and find 8 of them, which anticommute with
$\boldsymbol{n}\cdot\boldsymbol{\alpha}$. Thus we get
$$
K=c_{1}{\alpha}^{1} + c_{2}{\alpha}^{2} + {\rm
i}c_{3}{\alpha}^{1}(\boldsymbol{n}\cdot\boldsymbol{\alpha}) + {\rm
i}c_{4}{\alpha}^{2}(\boldsymbol{n}\cdot\boldsymbol{\alpha})
$$
$$
 + c_{5}\beta+{\rm
i}c_{6}\beta\gamma^{5}+{\rm
i}c_{7}\beta(\boldsymbol{n}\cdot\boldsymbol{\alpha})+c_{8}\beta(\boldsymbol{n}\cdot\boldsymbol{\alpha})\gamma^{5},\eqno(29)
$$
where coefficients $c_{j}\;(j=\overline{1,8})$ are real, since $K$
is Hermitian, and, as a consequence of (9), obey condition
$$
\frac{c_{1}}{c_{3}}=\frac{c_{2}}{c_{4}}=\frac{c_{5}}{c_{7}}=-\frac{c_{6}}{c_{8}}. \eqno(30)
$$
Defining parameters $\varphi$ and $\tilde{\varphi}$
by arranging terms in (29) into combinations $\exp({\rm i}\varphi\gamma^{5})$ 
and $\exp({\rm i}\tilde{\varphi}\boldsymbol{n}\cdot\boldsymbol{\alpha})$, where $ -\pi/2 < \varphi \leq \pi/2$ 
and $-\pi/2 \leq \tilde{\varphi} < \pi/2$, we
recast (29) into the form
$$
K=(\tilde{c}_{1}{\alpha}^{1} + \tilde{c}_{2}{\alpha}^{2} + \tilde{c}_{3}\beta{\rm
e}^{{\rm
i}\varphi\gamma^{5}}){\rm
e}^{{\rm
i}\tilde{\varphi}\boldsymbol{n}\cdot\boldsymbol{\alpha}}, \eqno(31)
$$
where real coefficients  $\tilde{c}_{j} \; (j=\overline{1,3})$, in view of (9), obey condition
$$
\tilde{c}_{1}^{2}+\tilde{c}_{2}^{2}+\tilde{c}_{3}^{2}=1. 
$$
With the use of obvious parametrization
$$
\tilde{c}_{1}=\sin\theta\cos\phi, \quad
\tilde{c}_{2}=\sin\theta\sin\phi, \quad
\tilde{c}_{3}=\cos\theta, 
$$
we finally obtain matrix $K$ in the form
$$
K=[\beta{\rm
e}^{{\rm
i}\varphi\gamma^{5}}\cos\theta+(\alpha^{1}\cos\phi+\alpha^{2}\sin\phi)\sin\theta]{\rm
e}^{{\rm
i}\tilde{\varphi}\boldsymbol{n}\cdot\boldsymbol{\alpha}}, \eqno(32)
$$
interpolating continuously (and smoothly) between $K_{+}$ and
$K_{-}$:
$$
K_{+}=K|_{\varphi=0, \, \tilde{\varphi}=-\pi/2, \, \theta=0}, \quad
K_{-}=K|_{\varphi=\pi/2, \, \tilde{\varphi}=0}.\eqno(33)
$$
The explicit form of the boundary condition ensuring the
self-adjointness of operator $H$ (4) in this case is
$$
\left\{I - [\beta{\rm
e}^{{\rm
i}\varphi\gamma^{5}}\cos\theta+(\alpha^{1}\cos\phi+\alpha^{2}\sin\phi)\sin\theta]{\rm
e}^{{\rm
i}\tilde{\varphi}\boldsymbol{n}\cdot\boldsymbol{\alpha}}\right\}\chi|_{\mathbf{r}\in
\partial{\Omega}}=0 \eqno(34)
$$
(the same condition is for $\tilde{\chi}$), and relation (11) takes
form
$$
\tilde{\chi}^{\dag}[\beta{\rm
e}^{{\rm
i}\varphi\gamma^{5}}\cos\theta+(\alpha^{1}\cos\phi+\alpha^{2}\sin\phi)\sin\theta]{\rm
e}^{{\rm
i}(\tilde{\varphi} + \pi/2)\boldsymbol{n}\cdot\boldsymbol{\alpha}}\chi|_{\mathbf{r}\in
\partial{\Omega}}=0.\eqno(35)
$$
Four parameters of boundary condition (34), $\varphi, \tilde{\varphi}, \theta$ and $\phi$, which 
vary arbitrarily from point to point of the boundary, are interpreted as the self-adjoint extension parameters.

In the context of the Casimir effect, one usually considers
spatial region $\Omega$ with a disconnected boundary consisting of
two connected components, $\partial{\Omega} =
\partial{\Omega}^{(+)}\bigcup\partial{\Omega}^{(-)}$. Choosing
coordinates $\mathbf{r}=(x,y,z)$ in such a way that $x$ and $y$
are tangential to the boundary, while $z$ is normal to it, we
identify the position of $\partial{\Omega}^{(\pm)}$ with, say,
$z=\pm{a/2}$. In general, there are 8 self-adjoint extension parameters: $\varphi_+$, $\tilde{\varphi}_+$, $\theta_+$  and
$\phi_+$ corresponding to $\partial{\Omega}^{(+)}$ and $\varphi_-$, $\tilde{\varphi}_-$, $\theta_-$  and $\phi_-$ corresponding to
$\partial{\Omega}^{(-)}$. However, if some symmetry is present,
then the number of self-adjoint extension parameters is diminished. For instance, if the boundary consists of two parallel planes, then the
cases differing by the values of $\phi_+$ or $\phi_-$ are physically indistinguishable, since they are related by a rotation around a normal
to the boundary.  To avoid this unphysical degeneracy, one has to fix
$$
\theta_+ = \theta_- = 0, \eqno(36)
$$
and there remains 4 self-adjoint extension parameters: $\varphi_+$, $\tilde{\varphi}_+$, $\varphi_-$ and $\tilde{\varphi}_-$. 
Operator $H$ (4) acting on functions which are defined in the region bounded by two parallel planes is self-adjoint, if the 
following condition holds:
$$
\left\{I - \beta\exp[{\rm i}(\varphi_{\pm}\gamma^{5} \pm
\tilde{\varphi}_{\pm}\alpha^{z})]\right\}\chi|_{z=\pm{a/2}}=0
\eqno(37)
$$
(the same condition holds for $\tilde{\chi}$). The latter ensures
the fulfilment of constraints
$$
\tilde{\chi}^{\dag}\alpha^{z}\chi|_{z=\pm{a/2}}=0 \eqno(38)
$$
and
$$
\tilde{\chi}^{\dag}\beta\exp\left\{{\rm i}[\varphi_{\pm}\gamma^{5}
\pm (\tilde{\varphi}_{\pm}
+\pi/2)\alpha^{z}]\right\}\chi|_{z=\pm{a/2}}=0.\eqno (39)
$$

It should be noted that, if one chooses the $K$-matrix to be non-Hermitian and off-diagonal in the standard representation, 
see (14), then, employing parametrization (17), the self-adjointness is implemented in the context of the Casimir effect 
with the use of boundary condition
$$
[I-\frac{\beta(\cosh^2\tilde{\vartheta}_{\pm}+1)-I\sinh^2\tilde{\vartheta}_{\pm}}{2{\rm
i}\cosh\tilde{\vartheta}_{\pm}}({\pm}{\alpha}^z\cosh\vartheta_{\pm}+\beta\gamma^{5}\sinh\vartheta_{\pm})]\chi|_{z=\pm{a/2}}=0
\eqno(40)
$$
(the same condition is for $\tilde{\chi}$), while the analogue of (39) takes form
$$
\frac{1}{2}\tilde{\chi}^{\dag}[I(\cosh^2\tilde{\vartheta}_{\pm}+1)+\beta\sinh^2\tilde{\vartheta}_{\pm}]
(\beta\cosh\vartheta_{\pm} \mp
{\alpha}^z\gamma^{5}\sinh\vartheta_{\pm})\chi|_{z=\pm{a/2}}=0.
\eqno(41)
$$

    \section{Induced vacuum energy in the magnetic field
    background}

The operator of a spinor field which is quantized in an ultrastatic
background is presented in the form
$$\hat{\Psi}(t,\mathbf{r})=\sum\!\!\!\!\!\!\!\!\!\!\!\int\limits_{E_{\lambda}>0}{\rm e}^{-{\rm i}E_{\lambda}t}\psi_{\lambda}(\mathbf{r})\hat{a}_{\lambda}
+\sum\!\!\!\!\!\!\!\!\!\!\!\int\limits_{E_{\lambda}<0}{\rm
e}^{-{\rm
i}E_{\lambda}t}\psi_{\lambda}(\mathbf{r})\hat{b}^{\dag}_{\lambda},\eqno(42)
$$
where
$\hat{a}^{\dag}_{\lambda}$ and $\hat{a}_{\lambda}$
($\hat{b}^{\dag}_{\lambda}$ and $\hat{b}_{\lambda}$) are the
spinor particle (antiparticle) creation and destruction operators,
satisfying anticommutation relations
$[\hat{a}_\lambda,\hat{a}_{\lambda'}^\dagger]_+=[\hat{b}_\lambda,\hat{b}_{\lambda'}^\dagger]_+=\left\langle
\lambda|\lambda'\right\rangle$,
wave functions $\psi_{\lambda}(\textbf{r})$ form a
complete set of solutions to the stationary Dirac equation
$$
H\psi_{\lambda}(\mathbf{r})=E_{\lambda}\psi_{\lambda}(\mathbf{r});\eqno(43)
$$
$\lambda$ is the set of parameters (quantum numbers) specifying a
one-particle state with energy $E_{\lambda}$; symbol
$\sum\!\!\!\!\!\!\!\int\,$ denotes summation over discrete and
integration (with a certain measure) over continuous values of
$\lambda$. Ground state $|\texttt{vac}>$ is defined by condition
$\hat{a}_\lambda|\texttt{vac}>=\hat{b}_\lambda|\texttt{vac}>=0$.
The temporal component of the operator of the energy-momentum tensor
is given by expression
$$
\hat{T}^{00}=\frac{\rm{i}}{4}[\hat{\Psi}^{\dag}({\partial_0}\hat{\Psi})-({\partial_0}\hat{\Psi}^{T})\hat{\Psi}^{{\dag}T}
-({\partial_0}\hat{\Psi}^{\dag})\hat{\Psi}+\hat{\Psi}^{T}({\partial_0}\hat{\Psi}^{{\dag}T})],\eqno(44)
$$
where superscript $T$ denotes a transposed spinor. Consequently, the formal expression for the vacuum expectation value of the energy
density is
$$
\varepsilon=<\texttt{vac}|\hat{T}^{00}|\texttt{vac}>=
-\frac{1}{2}\sum\!\!\!\!\!\!\!\!\!\int \,|E_{\lambda}|\psi_{\lambda}^{\dag}(\textbf{r})\psi_{\lambda}(\textbf{r}).\eqno(45)
$$

Let us consider the quantized charged massive spinor field
in the background of a static uniform magnetic field; then $\bnabla=\boldsymbol{\partial}-{\rm
i}e\mathbf{A}$
and the connection can be chosen as $\mathbf{A}=(-yB,0,0)$,
where $B$ is the value of the magnetic field strength which is directed along the
$z$-axis in Cartesian coordinates $\mathbf{r}=(x,y,z)$, $\mathbf{B}=(0,0,B)$. The
one-particle energy spectrum is
$$
E_{nk}=\pm\omega_{nk},\eqno(46)
$$
where
$$
\omega_{nk}=\sqrt{2n|eB|+k^{2}+m^{2}},\;-\infty<k<\infty,\;n=0,1,2,...\,
,\eqno(47)
$$
$k$ is the value of the wave number vector along the $z$-axis, and
$n$ labels the Landau levels. Using the explicit form of the complete set of solutions to the Dirac equation, see Appendix A, 
one can get that expression (45) takes form
$$
\varepsilon^{\infty}=-\frac{|eB|}{2\pi^{2}}\int\limits_{-\infty}^{\infty}{\rm d}k
\sum\limits_{n=0}^{\infty}\iota_{n}\omega_{nk},\eqno(48)
$$
where $\iota_{n}=1-\frac{1}{2} \delta_{n0}$; the superscript on the
left-hand side indicates that the magnetic field fills the whole
(infinite) space. The integral and the sum in (48) are divergent and
require regularization and renormalization. This problem has been
solved long ago by Heisenberg and Euler \cite{Hei2} (see also
\cite{Schw}), and we just list here their result
$$
\varepsilon^{\infty}_{\rm ren}=\frac{1}{8\pi^{2}}\int\limits_{0}^{\infty}\frac{{\rm d}\tau}{\tau}
{\rm e}^{-\tau}\left[\frac{eBm^{2}}{\tau}\coth\left(\frac{eB\tau}{m^{2}}\right)-\frac{m^{4}}{\tau^2}
-\frac{1}{3}e^{2}B^{2}\right];\eqno(49)
$$
note that the renormalization procedure includes subtraction at
$B=0$ and renormalization of the charge.

Let us turn now to the quantized charged massive spinor field in
the background of a static uniform magnetic field in spatial
region $\Omega$ bounded by two parallel planes
$\partial{\Omega}^{(+)}$ and $\partial{\Omega}^{(-)}$; the
position of $\partial{\Omega}^{(\pm)}$ is identified with
$z=\pm{a/2}$, and the magnetic field is orthogonal to the
boundary. The solution to (43) in region $\Omega$ is chosen as a
superposition of two plane waves propagating in opposite directions along the $z$-axis,
$$
\psi_{qnl}(\mathbf{r})=\psi_{qnk_{l}}(\mathbf{r})+\psi_{qn-k_{l}}(\mathbf{r}),\eqno(50)
$$
where the explicit form of $\psi_{qnk}(\mathbf{r})$ and  $\psi_{qn-k}(\mathbf{r})$ is given in Appendix A, 
and all restrictions on the values of coefficients $C_{j}$ and
$\tilde{C}_{j} \; (j=0,1,2)$ are withdrawn for a while. The
values of wave number vector $k_{l} \; (l=0,\pm1,\pm2, ... )$ are
determined from the boundary condition, see (37):
$$
\left\{I-\beta\exp[{\rm i}(\varphi_{\pm}\gamma^{5} \pm
\tilde{\varphi}_{\pm}\alpha^{z})]\right\}\psi_{qnl}(\mathbf{r})|_{z=\pm{a/2}}=0,
\quad  n\geq1 \eqno(51)
$$
and
$$
\biggl[I+\frac{\beta}{2}\biggl(\pm\alpha^{z}\gamma^{5}-1\biggr){\rm e}^{{\rm
i}(\varphi_{\pm}-\tilde\varphi_{\pm})\gamma^{5}}\Theta(\pm eB)
$$
$$
-\frac{\beta}{2}\biggl(\pm\alpha^{z}\gamma^{5}+1\biggr){\rm e}^{{\rm
i}(\varphi_{\pm}+\tilde\varphi_{\pm})\gamma^{5}}\Theta(\mp
eB)\biggr]\psi^{(0)}_{q0l}(\mathbf{r})|_{z=\pm{a/2}} =0, \eqno(52)
$$
where the step function is introduced as $\Theta(u)=1$ at $u>0$
and $\Theta(u)=0$ at $u<0$. The latter conditions can be rewritten
as conditions on the coefficients:
$$
\left\{\begin{array}{l}M_{11}^{(n)}C_{1}+M_{12}^{(n)}C_{2}+M_{13}^{(n)}\tilde{C}_{1}+M_{14}^{(n)}\tilde{C}_{2}=0
\\
%[3 mm]
M_{21}^{(n)}C_{1}+M_{22}^{(n)}C_{2}+M_{23}^{(n)}\tilde{C}_{1}+M_{24}^{(n)}\tilde{C}_{2}=0
\\
%[3 mm]
M_{31}^{(n)}C_{1}+M_{32}^{(n)}C_{2}+M_{33}^{(n)}\tilde{C}_{1}+M_{34}^{(n)}\tilde{C}_{2}=0
\\
%[3 mm]
M_{41}^{(n)}C_{1}+M_{42}^{(n)}C_{2}+M_{43}^{(n)}\tilde{C}_{1}+M_{44}^{(n)}\tilde{C}_{2}=0\end{array}
\right\}, \quad  n\geq1 \eqno(53)
$$
and
$$
\left\{\begin{array}{l}M_{11}^{(0)}C_{0}+M_{12}^{(0)}\tilde{C}_{0}=0 \\
%[3 mm]
M_{21}^{(0)}C_{0}+M_{22}^{(0)}\tilde{C}_{0}=0\end{array}
\right\},\eqno(54)
$$
where
%\newpage
$$
\left\{\begin{array}{l}M_{11}^{(n)}=\left\{(\omega_{nl}+m)\sin\left[{\rm sgn}(E_{nl})
\frac{\varphi_{{\rm sgn}(eB)}-\tilde{\varphi}_{{\rm sgn}(eB)}}{2}+\Theta(-E_{nl})\frac{\pi}{2}\right]\right. \\
\left. +{\rm i}k_{l}\cos\left[{\rm sgn}(E_{nl})
\frac{\varphi_{{\rm sgn}(eB)}-\tilde{\varphi}_{{\rm sgn}(eB)}}{2}+\Theta(-E_{nl})\frac{\pi}{2}\right]\right\}{\rm e}^{{\rm i}k_{l}a/2}, \\
[3 mm]
M_{12}^{(n)}={\rm i}\sqrt{2n|eB|}\cos\left[{\rm sgn}(E_{nl})
\frac{\varphi_{{\rm sgn}(eB)}-\tilde{\varphi}_{{\rm sgn}(eB)}}{2}+\Theta(-E_{nl})\frac{\pi}{2}\right]{\rm
e}^{{\rm i}k_{l}a/2}, \\
[3 mm]
M_{13}^{(n)}=M_{11}^{(n)*},M_{14}^{(n)}=-M_{12}^{(n)*}, \\
[3 mm]
M_{21}^{(n)}=-{\rm i}\sqrt{2n|eB|}\sin\left[{\rm sgn}(E_{nl})
\frac{\varphi_{{\rm sgn}(eB)}+\tilde{\varphi}_{{\rm sgn}(eB)}}{2}-\Theta(E_{nl})\frac{\pi}{2}\right]{\rm
e}^{{\rm i}k_{l}a/2}, \\
[3 mm]
M_{22}^{(n)}=\left\{(\omega_{nl}+m)\cos\left[{\rm sgn}(E_{nl})
\frac{\varphi_{{\rm sgn}(eB)}+\tilde{\varphi}_{{\rm sgn}(eB)}}{2}-\Theta(E_{nl})\frac{\pi}{2}\right]\right. \\
\left. +{\rm i}k_{l}\sin\left[{\rm sgn}(E_{nl})
\frac{\varphi_{{\rm sgn}(eB)}+\tilde{\varphi}_{{\rm sgn}(eB)}}{2}-\Theta(E_{nl})\frac{\pi}{2}\right]\right\}{\rm
e}^{{\rm i}k_{l}a/2}, \\
%[3 mm]
M_{23}^{(n)}=-M_{21}^{(n)*},M_{24}^{(n)}=M_{22}^{(n)*}, \\
[3 mm]
M_{31}^{(n)}=M_{33}^{(n)*},M_{32}^{(n)}=-M_{34}^{(n)*}, \\
[3 mm]
M_{33}^{(n)}=\left\{(\omega_{nl}+m)\cos\left[{\rm sgn}(E_{nl})
\frac{\varphi_{-{\rm sgn}(eB)}+\tilde{\varphi}_{-{\rm sgn}(eB)}}{2}-\Theta(E_{nl})\frac{\pi}{2}\right]\right. \\
\left. +{\rm i}k_{l}\sin\left[{\rm sgn}(E_{nl})
\frac{\varphi_{-{\rm sgn}(eB)}+\tilde{\varphi}_{-{\rm sgn}(eB)}}{2}-\Theta(E_{nl})\frac{\pi}{2}\right]\right\}{\rm e}^{{\rm
i}k_{l}a/2}, \\
[3 mm]
M_{34}^{(n)}=-{\rm
i}\sqrt{2n|eB|}\sin\left[{\rm sgn}(E_{nl})
\frac{\varphi_{-{\rm sgn}(eB)}+\tilde{\varphi}_{-{\rm sgn}(eB)}}{2}-\Theta(E_{nl})\frac{\pi}{2}\right]{\rm e}^{{\rm i}k_{l}a/2}, \\
[3 mm]
M_{41}^{(n)}=-M_{43}^{(n)*},M_{42}^{(n)}=M_{44}^{(n)*}, \\
[3 mm]
M_{43}^{(n)}={\rm i}\sqrt{2n|eB|}\cos\left[{\rm sgn}(E_{nl})
\frac{\varphi_{-{\rm sgn}(eB)}-\tilde{\varphi}_{-{\rm sgn}(eB)}}{2}+\Theta(-E_{nl})\frac{\pi}{2}\right]{\rm e}^{{\rm
i}k_{l}a/2}, \\
[3 mm]
M_{44}^{(n)}=\left\{(\omega_{nl}+m)\sin\left[{\rm sgn}(E_{nl})
\frac{\varphi_{-{\rm sgn}(eB)}-\tilde{\varphi}_{-{\rm sgn}(eB)}}{2}+\Theta(-E_{nl})\frac{\pi}{2}\right] \right.\\
\left. +{\rm i}k_{l}\cos\left[{\rm sgn}(E_{nl})
\frac{\varphi_{-{\rm sgn}(eB)}-\tilde{\varphi}_{-{\rm sgn}(eB)}}{2}+\Theta(-E_{nl})\frac{\pi}{2}\right]\right\}
{\rm e}^{{\rm i}k_{l}a/2}\end{array}\right\},
$$
$$
n\geq1\eqno(55)
$$
and
$$
\left\{\begin{array}{l}M_{11}^{(0)}=\left\{(\omega_{0l}+m)\sin\left[{\rm sgn}(E_{0l})
\frac{\varphi_{{\rm sgn}(eB)}-\tilde{\varphi}_{{\rm sgn}(eB)}}{2}+\Theta(-E_{0l})\frac{\pi}{2}\right]\right. \\
\left. +{\rm i}k_{l}\cos\left[{\rm sgn}(E_{0l})
\frac{\varphi_{{\rm sgn}(eB)}-\tilde{\varphi}_{{\rm sgn}(eB)}}{2}+\Theta(-E_{0l})\frac{\pi}{2}\right]\right\}
{\rm e}^{{\rm i}k_{l}a/2}, \\
[3 mm]
M_{12}^{(0)}=M_{11}^{(0)*},
M_{21}^{(0)}=M_{22}^{(0)*},\\
[3 mm]
M_{22}^{(0)}=\left\{(\omega_{0l}+m)\cos\left[{\rm sgn}(E_{0l})
\frac{\varphi_{-{\rm sgn}(eB)}+\tilde{\varphi}_{-{\rm sgn}(eB)}}{2}-\Theta(E_{0l})\frac{\pi}{2}\right]\right. \\
\left. +{\rm i}k_{l}\sin\left[{\rm sgn}(E_{0l})
\frac{\varphi_{-{\rm sgn}(eB)}+\tilde{\varphi}_{-{\rm sgn}(eB)}}{2}-\Theta(E_{0l})\frac{\pi}{2}\right]\right\}
{\rm e}^{{\rm i}k_{l}a/2}\end{array}\right\};\eqno(56)
$$
%\newpage
here, ${\rm sgn}(u)=\Theta(u)-\Theta(-u)$ is the sign function, 
we have chosen
$$
{\alpha}^{z}=\begin{pmatrix}
0&{\sigma}^{3}\\
{\sigma}^{3}&0
\end{pmatrix}
$$
and introduced notations
$$
\omega_{nl}\equiv\omega_{nk_{l}}=\sqrt{2n|eB|+k_{l}^{2}+m^{2}}\eqno(57)
$$
and, similarly, $E_{nl} \equiv E_{nk_{l}}$.

Thus, the spectrum of wave number vector $k_{l}$ is determined
from condition
$$
\det M^{(n)}=0,\eqno(58)
$$
where
\newpage
$$
\det M^{(n)}=(m+\omega_{nl})^{2}
$$
$$
\times\biggl\{{\rm e}^{2{\rm
i}k_{l}a}\biggl[m\cos\varphi_+ -
\omega_{nl}\,{\rm sgn}(E_{nl})\cos\tilde{\varphi}_+ -{\rm i}k_{l}\sin\tilde{\varphi}_+\biggr]
$$
$$
\times\biggl[m\cos\varphi_{-} -\omega_{nl}\,{\rm sgn}(E_{nl})\cos\tilde{\varphi}_{-} -
{\rm i}k_{l}\sin\tilde{\varphi}_{-}\biggr]
$$
$$
-2\biggl[m\cos\varphi_{+} -
\omega_{nl}\,{\rm sgn}(E_{nl})\cos\tilde{\varphi}_{+}\biggr]
\biggl[m\cos\varphi_{-} -\omega_{nl}\,{\rm sgn}(E_{nl})\cos\tilde{\varphi}_{-}\biggr]
$$
$$
- 2k_{l}^{2}\biggl[\cos(\varphi_{+}-\varphi_{-})-\cos\tilde{\varphi}_{+}\cos\tilde{\varphi}_{-}\biggr]
$$
$$
+{\rm e}^{-2{\rm i}k_{l}a}\biggl[m\cos\varphi_{+} -\omega_{nl}\,{\rm
sgn}(E_{nl})\cos\tilde{\varphi}_+ +{\rm
i}k_{l}\sin\tilde{\varphi}_+\biggr]
$$
$$
\times\biggl[m\cos\varphi_{-} -\omega_{nl}\,{\rm sgn}(E_{nl})\cos\tilde{\varphi}_{-} 
+{\rm i}k_{l}\sin\tilde{\varphi}_{-}\biggr]\biggr\}, \quad  n\geq1
\eqno(59)
$$
and
$$
\det M^{(0)}=
\cos\left[{\rm sgn}(E_{0l})\frac{\varphi_{{\rm sgn}(eB)}-\tilde{\varphi}_{{\rm sgn}(eB)}}{2}+\Theta(-E_{0l})\frac{\pi}{2}\right]
$$
$$
\times\sin\left[{\rm sgn}(E_{0l})\frac{\varphi_{-{\rm sgn}(eB)}+\tilde{\varphi}_{-{\rm sgn}(eB)}}{2}-\Theta(E_{0l})\frac{\pi}{2}\right]
$$
$$
\times\biggl\{{\rm e}^{{\rm
i}k_{l}a}\biggl[\biggl(m+\omega_{0l}\biggr)\tan\left({\rm
sgn}(E_{0l})\frac{\varphi_{{\rm sgn}(eB)}-\tilde{\varphi}_{{\rm
sgn}(eB)}}{2}+\Theta(-E_{0l})\frac{\pi}{2}\right)+{\rm
i}k_{l}\biggr]
$$
$$
\times\biggl[\biggl(m+\omega_{0l}\biggr)\cot\left({\rm
sgn}(E_{0l})\frac{\varphi_{-{\rm sgn}(eB)}+\tilde{\varphi}_{-{\rm
sgn}(eB)}}{2}-\Theta(E_{0l})\frac{\pi}{2}\right)+{\rm
i}k_{l}\biggr]
$$
$$
- {\rm e}^{-{\rm i}k_{l}a}\biggl[\biggl(m+\omega_{0l}\biggr)
\tan\left({\rm sgn}(E_{0l})\frac{\varphi_{{\rm
sgn}(eB)}-\tilde{\varphi}_{{\rm
sgn}(eB)}}{2}+\Theta(-E_{0l})\frac{\pi}{2}\right)-{\rm
i}k_{l}\biggr]$$
$$
\times\biggl[\biggl(m+\omega_{0l}\biggr)\cot\left({\rm
sgn}(E_{0l})\frac{\varphi_{-{\rm sgn}(eB)}+\tilde{\varphi}_{-{\rm
sgn}(eB)}}{2}-\Theta(E_{0l})\frac{\pi}{2}\right)-{\rm
i}k_{l}\biggr]\biggr\}.\eqno(60)
$$

Given solution $\psi_{q0l}^{(0)}(\mathbf{r})$, we impose the condition on its
coefficients $C_{0}$ and $\tilde{C}_{0}$:
$$
\left\{\begin{array}{l}|C_{0}|^{2}+|\tilde{C}_{0}|^{2}=\frac{2\pi}{a},\\
[3 mm] C_{0}^{*} \tilde{C}_{0}+\tilde{C}_{0}^{*}
C_{0}=0;\end{array}\right.\eqno(61)
$$
in particular, the coefficients can be chosen as
$$
C_{0}=\sqrt{\frac{\pi}{a}}{\rm e}^{\rm{i}\pi/4},\quad
\tilde{C}_{0}=\sqrt{\frac{\pi}{a}}{\rm e}^{-\rm{i}\pi/4}.
$$
In the case of $n\geq1$, two linearly independent solutions, $\psi_{qnl}^{(1)}(\mathbf{r})$ and $\psi_{qnl}^{(2)}(\mathbf{r})$, are
orthogonal, if the appropriate coefficients, $C_{j}^{(1)},\,
\tilde{C}_{j}^{(1)}$ and $C_{j}^{(2)},\, \tilde{C}_{j}^{(2)}$
$(j=1,2)$, obey condition
$$
\left\{\begin{array}{l}\sum\limits_{j=1,2}C_{j}^{(1)*}C_{j}^{(2)}=0,\\
[3 mm]
C_{j}^{(1)}C_{j'}^{(2)}=\tilde{C}_{j}^{(1)}\tilde{C}_{j'}^{(2)},\,
|C_{j}^{(j')}|=|\tilde{C}_{j}^{(j')}|,\quad
j,j'=1,2.\end{array}\right.\eqno(62)
$$
We impose further condition:
$$
%\begin{displaymath}
\left\{\begin{array}{l}\sum\limits_{j=1,2}|C_{j}^{(j')}|^{2}=\frac{\pi}{a},\\
[6 mm]
\sum\limits_{j=1,2} [C_{j}^{(j')*} \tilde{C}_{j}^{(j')} + \tilde{C}_{j}^{(j')*}C_{j}^{(j')}]=0,\quad
j'=1,2;\end{array}\right.\eqno(63)
$$
in particular, the coefficients can be chosen as
$$
C_{1}^{(1)}=\sqrt{\frac{\pi}{2a}}{\rm
e}^{\rm{i}\pi/4},\,\tilde{C}_{1}^{(1)}=\sqrt{\frac{\pi}{2a}}{\rm
e}^{-\rm{i}\pi/4},\,C_{2}^{(1)}=\sqrt{\frac{\pi}{2a}}{\rm
e}^{-\rm{i}\pi/4},\,\tilde{C}_{2}^{(1)}=\sqrt{\frac{\pi}{2a}}{\rm
e}^{-3\rm{i}\pi/4}
$$
and
$$
C_{1}^{(2)}=\sqrt{\frac{\pi}{2a}}{\rm
e}^{-\rm{i}\pi/4},\,\tilde{C}_{1}^{(2)}=\sqrt{\frac{\pi}{2a}}{\rm
e}^{\rm{i}\pi/4},\,C_{2}^{(2)}=\sqrt{\frac{\pi}{2a}}{\rm
e}^{\rm{i}\pi/4},\,\tilde{C}_{2}^{(2)}=\sqrt{\frac{\pi}{2a}}{\rm
e}^{3\rm{i}\pi/4}.
$$
As a result, wave functions
$\psi^{(j)}_{qnl}(\mathbf{r})\,(j=0,1,2)$ satisfy the requirements
of orthonormality
$$\int\limits_{\Omega}{{\rm d}^{3}r}\,{\psi^{(j)\dag}_{qnl}(\mathbf{r})}\psi_{q'n'l'}^{(j')}(\mathbf{r})=
\delta_{jj'}\delta_{nn'}\delta_{ll'}\delta(q-q'), \quad j,j'=0,1,2\eqno(64)
$$
and completeness
$$
\sum\limits_{{\rm sgn}(E_{nl})}\int\limits_{-\infty}^{\infty}{\rm{d}}q\sum\limits_{l}\left[\psi^{(0)}_{q0l}(\mathbf{r})\psi^{(0)\dag}_{q0l}(\mathbf{r'})
+\sum\limits_{n=1}^{\infty}\sum\limits_{j=1,2}\psi^{(j)}_{qnl}(\mathbf{r})\psi^{(j)\dag}_{qnl}(\mathbf{r'})\right]=
I\delta(\mathbf{r}-\mathbf{r'}).\eqno(65)
$$
Consequently, we obtain the following formal expression for the
vacuum expectation value of the energy per unit area of the boundary
surface
$$
\frac{E}{S}\equiv\int\limits_{-a/2}^{a/2}{\rm{d}}z\,\varepsilon=
-\frac{|eB|}{2\pi}\sum\limits_{{\rm sgn}(E_{nl})}\sum\limits_{l}\sum\limits_{n=0}^{\infty}\iota_{n}\omega_{nl}.\eqno(66)
$$

Concluding this section we recall that, owing to the boundary condition, see (51)
and (52), the normal component of current
$\boldsymbol{J}_{qnlj}(\mathbf{r})=\psi_{qnl}^{(j)\dag}(\mathbf{r})\boldsymbol{\alpha}\psi_{qnl}^{(j)}(\mathbf{r})$ 
$(j=0,1,2)$ vanishes at the boundary, see (38),
$$
J^{z}_{qnlj}(\mathbf{r})|_{z=\pm{a/2}}=0, \eqno(67)
$$
which, cf. (1), signifies that the quantized matter is confined within the
boundaries.

\section{Choice of boundary conditions, Casimir\\ energy and force}

The spectrum of the wave number vector in the direction of the
magnetic field, which is determined from (58), depends on four self-adjoint extension parameters,
$\varphi_{+},\tilde{\varphi}_{+},\varphi_{-}$ and
$\tilde{\varphi}_{-}$, in the case of $n\geq1$, see (59), and on
two self-adjoint extension parameters, $\varphi_{+}-\tilde{\varphi}_{+}$ and
$\varphi_{-}+\tilde{\varphi}_{-}$  ($eB>0$), or $\varphi_{+}+\tilde{\varphi}_{+}$ and
$\varphi_{-}-\tilde{\varphi}_{-}$  ($eB<0$), in the case of $n=0$, see (60). As was mentioned in Section 2, 
the values of these self-adjoint extension parameters may
vary arbitrarily from point to point of the boundary surface.
However, in the context of the Casimir effect, such a generality seems to be excessive, lacking physical
motivation and, moreover, being impermissible, as long as boundary condition (51)-(52) is to be regarded as the 
one determining the spectrum of the wave number vector in the $z$-direction.
Therefore, we shall assume in the following that the self-adjoint
extension parameters are independent of coordinates $x$ and $y$.

The equation determining the spectrum of $k_{l}$, see (58), can be presented in the form
$$
{\rm e}^{2{\rm i}k_{l}a}={\rm e}^{-2{\rm i}\eta_{k_{l}}},\eqno(68)
$$
or
$$
\sin(k_{l}a+\eta_{k_{l}})=0,\eqno(69)
$$
where
$$
\eta_{k_l}=-\frac 12{\rm arctan}\left\{k_l\sin\tilde{\varphi}_+
\left[m\cos\varphi_+-\omega_{nl}\,{\rm
sgn}(E_{nl})\cos\tilde{\varphi}_+\right]^{-1}\right\}
$$
$$
-\frac 12{\rm
arctan}\left\{k_l\sin\tilde{\varphi}_-\left[(m\cos\varphi_--\omega_{nl}\,{\rm
sgn}(E_{nl})\cos\tilde{\varphi}_-\right]^{-1}\right\}
$$
$$
\mp\frac 12{\rm arctan}\Biggl(k_l\Biggl\{m^2\biggl[2-2\cos(\varphi_+-\varphi_-)
(\cos\varphi_+\cos\varphi_-+\cos\tilde{\varphi}_+\cos\tilde{\varphi}_-)
%\right.
$$
$$
\left.\left.
%\left.
+2\cos\varphi_+\cos\varphi_-\cos\tilde{\varphi}_+\cos\tilde{\varphi}_--\sin^2\varphi_+
\sin^2\tilde{\varphi}_- - \sin^2\varphi_-\sin^2\tilde{\varphi}_+\biggr]\right.\right.
$$
$$
\left.\left.-2m \omega_{nl}\,{\rm sgn}(E_{nl})\biggl[\cos\varphi_+
\cos\tilde{\varphi}_++\cos\varphi_-\cos\tilde{\varphi}_-\right.\right.
$$
$$
\left.\left.\left.-\cos(\varphi_+-\varphi_-)\left(\cos\varphi_+\cos\tilde{\varphi}_-
+\cos\varphi_-\cos\tilde{\varphi}_+\right)\biggr]\right.\right.
\right.
$$
$$
-2n|eB|\biggl[2\cos(\varphi_+-\varphi_-)\cos\tilde{\varphi}_+
\cos\tilde{\varphi}_--\cos^2\tilde{\varphi}_+-\cos^2\tilde{\varphi}_-\biggr]
$$
$$
\left.\left.+k^2_l\sin^2(\varphi_+-\varphi_-)\Biggr\}^{1/2}
\Biggl[m^2\left(\cos\varphi_+\cos\varphi_-+\cos\tilde{\varphi}_+\cos\tilde{\varphi}_-\right)
\right.\right.
$$
$$
\left.-m \omega_{nl}\,{\rm sgn}(E_{nl})\left(\cos\varphi_+\cos\tilde{\varphi}_-+\cos\varphi_-\cos\tilde{\varphi}_+\right)
+2n|eB|\cos\tilde{\varphi}_+\cos\tilde{\varphi}_-\right.
$$
$$
\left.\left.
+k^2_l\cos(\varphi_+-\varphi_-)\Biggr]^{-1}\Biggr),
\right. \quad  n\geq 1  \right. \eqno(70)
$$
(two signs correspond to two roots of the quadratic equation for
variable ${\rm e}^{2{\rm i}k_{l}a}$, see (58) and (59)) and
$$
\eta_{k_{l}}=\arctan\Biggl(k_{l}\sin\Biggl[\frac{1}{2}\left(\varphi_{{\rm
sgn}(eB)}-\varphi_{-{\rm sgn}(eB)}-\tilde{\varphi}_{{\rm
sgn}(eB)}-\tilde{\varphi}_{-{\rm sgn}(eB)}\right)\Biggr]
$$
$$
\times\Biggl\{m\cos\Biggl[\frac{1}{2}\left(\varphi_{{\rm
sgn}(eB)}+\varphi_{-{\rm sgn}(eB)}-\tilde{\varphi}_{{\rm
sgn}(eB)}+\tilde{\varphi}_{-{\rm sgn}(eB)}\right)\Biggr]
$$
$$
-\omega_{0l}{\rm
sgn}(E_{0l})\cos\Biggl[\frac{1}{2}\left(\varphi_{{\rm
sgn}(eB)}-\varphi_{-{\rm sgn}(eB)}-\tilde{\varphi}_{{\rm
sgn}(eB)}-\tilde{\varphi}_{-{\rm
sgn}(eB)}\right)\Biggr]\Biggr\}^{-1}\Biggr), \quad  n=0. \eqno(71)
$$

It should be emphasized that value $k_l=0$ is allowed for special cases only. Really,
we have in the case of $k_l=0$:
$$
\psi_{qnl}^{(j)}({\bf r})|_{z=a/2}=\psi_{qnl}^{(j)}({\bf
r})|_{z=-a/2}, \eqno(72)
$$
and boundary condition (51)-(52) can be presented in the form
$$
R \, \psi_{qnl}^{(j)}({\bf r})|_{k_l=0}=0, \eqno(73)
$$
where
$$ \left\{
\begin{array}{l}R_{11}=\sin\frac{\varphi_+-\tilde{\varphi}_+}{2}, \quad
R_{12}=0, \quad R_{13}={\rm i}\cos\frac{\varphi_+-\tilde{\varphi}_+}{2}, \quad R_{14}=0, \\
R_{21}=0, \quad R_{22}=\sin\frac{\varphi_++\tilde{\varphi}_+}{2}, \quad R_{23}=0, \quad 
R_{24}={\rm i}\cos\frac{\varphi_++\tilde{\varphi}_+}{2}, \\
R_{31}=\sin\frac{\varphi_-+\tilde{\varphi}_-}{2}, \quad R_{32}=0,
\quad R_{33}={\rm i}\cos\frac{\varphi_-+\tilde{\varphi}_-}{2}, \quad R_{34}=0,
\\
R_{41}=0, \quad R_{42}=\sin\frac{\varphi_--\tilde{\varphi}_-}{2}, \quad R_{43}=0,
\quad R_{44}={\rm i}\cos\frac{\varphi_--\tilde{\varphi}_-}{2}
\end{array}\right\}. \eqno(74)
$$
The determinant of matrix $R$ is:
$$
\det R = -\sin\frac{\varphi_+-\varphi_-+\tilde{\varphi}_++\tilde{\varphi}_-}{2}
\sin\frac{\varphi_+-\varphi_--\tilde{\varphi}_+-\tilde{\varphi}_-}{2}. \eqno(75)
$$
The necessary and sufficient condition for value $k_l=0$ to be admissible is $\det R = 0$, i.e. either 
$$
\varphi_+-\varphi_-=\tilde{\varphi}_++\tilde{\varphi}_-, \eqno(76)
$$
or 
$$
\varphi_+-\varphi_-=-\tilde{\varphi}_+-\tilde{\varphi}_-. \eqno(77)
$$
Otherwise, $\det R \neq 0$ and value $k_l=0$ is excluded from the spectrum, 
because equation (73) then allows for the trivial solution only, $\psi_{qnl}^{(j)}({\bf r})|_{k_l=0} \equiv 0$.

It is not clear which of the signs in (70) should be chosen.
Moreover, the square root in the argument of the last arctangent
in (70) may become imaginary for some values of the self-adjoint
extension parameters, and this results in the complex values of
$k_l$. Both of these obstructions are clearly eliminated by imposing restriction
$$
\varphi_{+}=\varphi_{-}=\varphi, \quad
\tilde{\varphi}_{+}=\tilde{\varphi}_{-}=\tilde{\varphi}. \eqno(78)
$$
Then (70) and (71) take form
$$
\eta_{k_{l}}=-\arctan\biggl[k_{l}\sin\tilde{\varphi}\biggl(m\cos\varphi-
{\rm sgn}(E_{nl})\omega_{nl}\cos\tilde{\varphi}\biggr)^{-1}\biggr], \quad  n\geq 0, \eqno(79)
$$
and the equation determining the spectrum of $k_{l}$ can be presented as
$$
\cos(k_{l}a)+\frac{\omega_{nl}\,{\rm sgn}(E_{nl})\cos\tilde{\varphi}-m\cos\varphi}{k_{l}\sin\tilde{\varphi}}\sin(k_{l}a)
=0; \eqno(80)
$$
note that the spectrum consists of values of the same sign, say, $k_l>0$ (values of
the opposite sign ($k_l<0$) should be excluded to avoid double
counting). Relations (37) and (39) in the case of (78) take forms
$$
\left\{I - \beta\exp[{\rm i}(\varphi\gamma^{5} \pm
\tilde{\varphi}\alpha^{z})]\right\}\chi|_{z=\pm{a/2}}=0
\eqno(81)
$$
and
$$
\tilde{\chi}^{\dag}\beta\exp\left\{{\rm i}[\varphi\gamma^{5}
\pm (\tilde{\varphi}
+\pi/2)\alpha^{z}]\right\}\chi|_{z=\pm{a/2}}=0.\eqno (82)
$$
respectively. 

In the case of $\tilde{\varphi}=- \pi/2$, the spectrum of $k_{l}$ is independent of the number of the Landau level, $n$, 
and of the sign of the one-particle energy, ${\rm sgn}(E_{nl})$; it is determined from equation  
$$
\cos(k_{l}a)+\frac{m\cos\varphi}{k_{l}}\sin(k_{l}a)=0. \eqno(83)
$$
In the case of $\tilde{\varphi}=0$, the $k_{l}$-spectrum is also independent of $n$ and of ${\rm sgn}(E_{nl})$; moreover, 
it is independent of $\varphi$, since the determining equation takes form
$$
\sin(k_{l}a)=0; \eqno(84)
$$
note that value $k_l=0$ is admissible in this case, see (76)-(78).
In what follows, we shall consider the most general case of two self-adjoint extension parameters, 
$\varphi$ and $\tilde{\varphi}$, when the $k_{l}$-spectrum depends on $n$ and on ${\rm sgn}(E_{nl})$, see (80).

As was already mentioned, the expression for the induced vacuum energy per unit area of the boundary
surface, see (66), can be regarded as purely formal, since it is
ill-defined due to the divergence of infinite sums over $l$ and $n$.
To tame the divergence, a factor containing a regularization
parameter should be inserted in (66). A summation over values $k_{l}\geq 0$, which are
determined by (80), can be performed with the use of the Abel-Plana formula and its generalizations. 
In the cases of $\tilde{\varphi}=0$ and of $\varphi=-\tilde\varphi=\pi/2$, the well-known versions of the Abel-Plana
formula (see, e.g., \cite{Bor}),
\newpage
$$
\left.\sum\limits_{{\rm sgn}(E_{nl})}\sum\limits_{k_{l}\geq 0}f(k_{l}^{2})\right|_{\sin(k_{l}a)=0}=
\frac{a}{\pi}\int\limits_{-\infty}^{\infty}{\rm{d}}k{f(k^{2})}-\frac{2{\rm
i}a}{\pi}\int\limits_{0}^{\infty}{\rm{d}}\kappa \frac{f[(-{\rm
i}\kappa)^{2}]-f[({\rm i}\kappa)^{2}]}{{\rm e}^{2{\kappa}a}-1}
$$
$$
+f(0) \eqno(85)
$$
and
$$
\left.\sum\limits_{{\rm sgn}(E_{nl})}\sum\limits_{k_{l}>0}f(k_{l}^{2})\right|_{\cos(k_{l}a)=0}=
\frac{a}{\pi}\int\limits_{-\infty}^{\infty}{\rm{d}}k{f(k^{2})}+\frac{2{\rm
i}a}{\pi}\int\limits_{0}^{\infty}{\rm{d}}\kappa \frac{f[(-{\rm
i}\kappa)^{2}]-f[({\rm i}\kappa)^{2}]}{{\rm e}^{2{\kappa}a}+1},
\eqno(86)
$$
are used, respectively. Otherwise, we use the version of the Abel-Plana formula,
that is derived in Appendix B:
$$
\sum\limits_{{\rm sgn}(E_{nl})}\sum\limits_{k_{l}>0}f(k_{l}^{2})=\frac{a}{\pi}\int\limits_{-\infty}^{\infty}{\rm{d}}k{f(k^{2})}
+\frac{2{\rm
i}a}{\pi}\int\limits_{0}^{\infty}{\rm{d}}\kappa\Lambda(\kappa)\{f[(-{\rm
i}\kappa)^{2}]-f[({\rm i}\kappa)^{2}]\}
$$
$$
-f(0)-\frac{1}{\pi}
\int\limits_{-\infty}^{\infty}{\rm{d}}k f(k^{2})
\frac{m\cos\varphi\sin\tilde\varphi[k^{2}-\mu_n(\varphi,\tilde\varphi)]}
{[k^{2}+\mu_n(\varphi,\tilde\varphi)]^2 +4k^{2}m^{2}\cos^{2}\varphi\sin^{2}\tilde\varphi},\eqno(87)
$$
where
$$
\Lambda(\kappa)=\Biggl( -[\kappa^2\cos2\tilde\varphi-\mu_n(\varphi,\tilde\varphi)]{\rm e}^{2{\kappa}a}
+{\kappa}^{2}+2{\kappa}m\cos\varphi\sin\tilde\varphi-\mu_n(\varphi,\tilde\varphi)
$$
$$
+\frac{\sin\tilde\varphi}{a}\left\{-\kappa^2 m\cos\varphi(\cos2\tilde\varphi{\rm e}^{2{\kappa}a}-1)
+[(2{\kappa}\sin\tilde\varphi-m\cos\varphi){\rm e}^{2{\kappa}a}\right.
$$
$$
\left.+m\cos\varphi]\mu_n(\varphi,\tilde\varphi)
\right\}[{\kappa}^{2}-2{\kappa}m\cos\varphi\sin\tilde\varphi-\mu_n(\varphi,\tilde\varphi)]^{-1}\Biggr)
$$
$$
\times
\left\{[{\kappa}^{2}-2{\kappa}m\cos\varphi\sin\tilde\varphi-\mu_n(\varphi,\tilde\varphi)]
{\rm e}^{4{\kappa}a}\right.
$$
$$
\left. -2[\kappa^2\cos2\tilde\varphi-\mu_n(\varphi,\tilde\varphi)]{\rm e}^{2{\kappa}a}
+{\kappa}^{2}+2{\kappa}m\cos\varphi\sin\tilde\varphi-\mu_n(\varphi,\tilde\varphi)\right\}^{-1} \eqno(88)
$$
and
$$
\mu_n(\varphi,\tilde\varphi)=2n|eB|\cos^{2}\tilde\varphi
+m^{2}\sin(\varphi+\tilde\varphi)\sin(\varphi-\tilde\varphi). \eqno(89)
$$
In (85)-(87), $f(u^{2})$ as a function of complex variable $u$ is assumed to decrease
sufficiently fast at large distances from the origin of the complex
$u$-plane, and this decrease is due to the use of some kind of regularization for (66). However, the regularization 
in the second integral on the right-hand side of (85)-(87) can be removed; then
$$
{\rm i}\{f[(-{\rm i}\kappa)^{2}]-f[({\rm
i}\kappa)^{2}]\}=-\frac{|eB|}{\pi}\sum\limits_{n=0}^{\infty}\iota_{n}\sqrt{\kappa^{2}-\omega^{2}_{n0}}
$$
with the range of $\kappa$ restricted to
$\kappa>\omega_{n0}$ for the corresponding terms; here, recalling (47),
$\omega_{n0}=\sqrt{2n|eB|+m^{2}}$.
As to the first integral on the right-hand side of (85)-(87), one immediately
recognizes that it is equal to $\varepsilon^{\infty}$ (48)
multiplied by $a$. Hence, if one ignores for a moment the last term
of (85), as well as the terms in the last line of (87), then the problem of regularization and removal
of the divergency in expression (66) is the same as that in the case
of no boundaries, when the magnetic field fills the whole space.
Defining the Casimir energy as the vacuum energy per unit area of the boundary surface,
which is renormalized in the same way as in the case of no
boundaries, we obtain
$$
\frac{E_{\rm ren}}{S}=a\varepsilon^{\infty}_{\rm ren}-
\frac{2|eB|}{\pi^{2}}a\sum\limits_{n=0}^{\infty}\iota_{n}\int\limits_{\omega_{n0}}^{\infty}{\rm
d}\kappa\Lambda(\kappa)\sqrt{\kappa^{2}-\omega^{2}_{n0}}
$$
$$
+\frac{|eB|}{2\pi}\sum\limits_{n=0}^{\infty}\iota_{n}\omega_{n0}+\frac{|eB|}{2\pi^{2}}\int\limits_{-\infty}^{\infty}{\rm
d}k
\sum\limits_{n=0}^{\infty}\iota_{n}\sqrt{k^{2}+\omega^{2}_{n0}}
$$
$$
\times\frac{m\cos\varphi\sin\tilde\varphi[k^{2}-\mu_n(\varphi,\tilde\varphi)]}
{[k^{2}+\mu_n(\varphi,\tilde\varphi)]^2 +4k^{2}m^{2}\cos^{2}\varphi\sin^{2}\tilde\varphi},\eqno(90)
$$
$\varepsilon^{\infty}_{\rm ren}$ is given by (49). The sums and
the integral in the last two lines of (90)
(which are due to the terms in the last line of (87) and which can be interpreted as describing the proper
energies of the boundary planes containing the sources of the
magnetic field) are divergent, but this divergency is
of no concern for us, because it has no physical consequences.
Rather than the Casimir energy, a physically relevant quantity is
the Casimir force per unit area of the boundary surface, i.e. pressure, which is defined as
$$
F=-\frac{\partial}{\partial a}\frac{E_{\rm ren}}{S},\eqno(91)
$$
and which is free from divergencies. We obtain
$$
F=-\varepsilon^{\infty}_{\rm ren}-
\frac{2|eB|}{\pi^{2}}\sum\limits_{n=0}^{\infty}\iota_{n}\int\limits_{\omega_{n0}}^{\infty}{\rm
d}\kappa\Upsilon(\kappa)\sqrt{\kappa^{2}-\omega_{n0}^{2}},\eqno(92)
$$
where
$$
\Upsilon(\kappa)\equiv-\frac{\partial}{\partial a}a\Lambda(\kappa)
=\left[\upsilon_{1}(\kappa){\rm e}^{6{\kappa}a}+\upsilon_{2}(\kappa){\rm e}^{4{\kappa}a}+
\upsilon_{3}(\kappa){\rm e}^{2{\kappa}a}+\upsilon_{4}(\kappa)\right]
$$
$$
\times\left\{[{\kappa}^{2}-2{\kappa}m\cos\varphi\sin\tilde\varphi-\mu_n(\varphi,\tilde\varphi)]
{\rm e}^{4{\kappa}a}\right.
$$
$$
\left. -2[\kappa^2\cos2\tilde\varphi-\mu_n(\varphi,\tilde\varphi)]{\rm e}^{2{\kappa}a}
+{\kappa}^{2}+2{\kappa}m\cos\varphi\sin\tilde\varphi-\mu_n(\varphi,\tilde\varphi)\right\}^{-2} \eqno(93)
$$
and
$$
\upsilon_{1}(\kappa)=-(2\kappa a-1)
[{\kappa}^{2}-2{\kappa}m\cos\varphi\sin\tilde\varphi-\mu_n(\varphi,\tilde\varphi)]
[\kappa^2\cos2\tilde\varphi-\mu_n(\varphi,\tilde\varphi)]
$$
$$
-2[{\kappa}^{2}m\cos\varphi\cos2\tilde\varphi
-(2{\kappa}\sin\tilde\varphi-m\cos\varphi)\mu_n(\varphi,\tilde\varphi)]\kappa\sin\tilde\varphi, \eqno(94)
$$
$$
\upsilon_{2}(\kappa)=(4\kappa a-3)
\left\{[{\kappa}^{2}-\mu_n(\varphi,\tilde\varphi)]^{2}
-4{\kappa}^{2}m^{2}\cos^{2}\varphi\sin^{2}\tilde\varphi\right\}
$$
$$
+8\kappa^2[\kappa^2\cos^{2}\tilde\varphi-m^{2}\cos^{2}\varphi-\mu_n(\varphi,\tilde\varphi)]\sin^{2}\tilde\varphi
%$$
%$$
+4[\kappa^2+\mu_n(\varphi,\tilde\varphi)]{\kappa}m\cos\varphi\sin\tilde\varphi, \eqno(95)
$$
$$
\upsilon_{3}(\kappa)=
-(2\kappa a-3)[{\kappa}^{2}+2{\kappa}m\cos\varphi\sin\tilde\varphi-\mu_n(\varphi,\tilde\varphi)]
[\kappa^2\cos2\tilde\varphi-\mu_n(\varphi,\tilde\varphi)]
$$
$$
-2[{\kappa}^{2}m\cos\varphi\cos2\tilde\varphi
+(2{\kappa}\sin\tilde\varphi+m\cos\varphi)\mu_n(\varphi,\tilde\varphi)]\kappa\sin\tilde\varphi, \eqno(96)
$$
$$
\upsilon_{4}(\kappa)=
-[{\kappa}^{2}+2{\kappa}m\cos\varphi\sin\tilde\varphi-\mu_n(\varphi,\tilde\varphi)]^{2}. \eqno(97) 
$$

\section{Some particular cases}

In the cases of $\tilde\varphi={-\pi}/{2}$ and of $\tilde\varphi=0$, relations (93)-(97) are simplified: 
$$
\left.\Upsilon(\kappa)\right|_{\tilde\varphi={-\pi}/{2}}
=\frac{\left[\left(2\kappa a-1\right)
\left(\kappa^{2}-m^{2}\cos^{2}\varphi\right)-2{\kappa}m\cos\varphi\right]{\rm
e}^{2{\kappa}a}-\left(\kappa-m\cos\varphi\right)^{2}}
{\left[\left(\kappa+m\cos\varphi\right){\rm
e}^{2{\kappa}a}+\kappa-m\cos\varphi\right]^{2}} \eqno(98)
$$
and
$$
\left.\Upsilon(\kappa)\right|_{\tilde\varphi=0}
=-\frac{\left(2\kappa a-1\right)
{\rm e}^{2{\kappa}a}+1}
{\left({\rm e}^{2{\kappa}a}-1\right)^{2}}. \eqno(99)
$$

Recalling the choice of off-diagonal matrix $K$ (14) at $u^2-v^2-{\boldsymbol{t}}^2 > 0$, see (18), we note that boundary 
condition (40) under restriction 
$$
\vartheta_{+}=\vartheta_{-}=\vartheta, \quad
\tilde{\vartheta}_{+}=\tilde{\vartheta}_{-}=0, \eqno(100)
$$
takes form
$$
(I \pm {\rm i}\beta{\alpha}^z\cosh\vartheta + {\rm
i}\gamma^{5}\sinh\vartheta)\chi|_{z=\pm{a/2}}=0. \eqno(101)
$$
As has been proven in \cite{Si1}, the appropriate equation determining the spectrum of $k_{l}$ in this case is
$$
\cos(k_{l}a)+\frac{m}{k_{l}\cosh\vartheta}\sin(k_{l}a)=0,
\eqno(102)
$$
and the Casimir pressure is given by (92) with 
$$
\Upsilon(\kappa)
=\frac{\left[\left(2\kappa a-1\right)
\left(\kappa^{2}\cosh^{2}\vartheta-m^{2}\right)-2{\kappa}m\cosh\vartheta\right]{\rm
e}^{2{\kappa}a}-\left(\kappa\cosh\vartheta-m\right)^{2}}
{\left[\left(\kappa\cosh\vartheta+m\right){\rm
e}^{2{\kappa}a}+\kappa\cosh\vartheta-m\right]^{2}}. \eqno(103)
$$
Comparing (83) and (98) with (102) and (103), we see that the case of the Hermitian $K$-matrix 
restricted by (78) and condition $\tilde\varphi={-\pi}/{2}$ is obtainable by substitution $\cosh\vartheta \rightarrow 1/\cos\varphi$ 
from the case of the off-diagonal $K$-matrix restricted by (100); 
the latter case was exhaustively studied in \cite{Si1}. We only remind here that the case of the MIT bag boundary condition 
corresponds to $\varphi=0$, or, respectively, $\vartheta=0$.

In the case of $\tilde\varphi=0$, we obtain
$k_{l}=\frac{\pi}{a}l \,\, (l=0,1,2,...)$ and
$$ 
\left.\frac{E_{\rm
ren}}{S}\right|_{\tilde\varphi=0}=a\varepsilon^{\infty}_{\rm
ren}+
\frac{2|eB|}{\pi^{2}}a\sum\limits_{n=0}^{\infty}\iota_{n}\int\limits_{\omega_{n0}}^{\infty}{\rm
d}\kappa\frac{\sqrt{\kappa^{2}-\omega_{n0}^{2}}}{{\rm
e}^{2{\kappa}a}-1}-\frac{|eB|}{2\pi}\sum\limits_{n=0}^{\infty}\iota_{n}\omega_{n0}\eqno(104)
$$
and
$$
\left.F\right|_{\tilde\varphi=0}=-\varepsilon^{\infty}_{\rm
ren}+
\frac{2|eB|}{\pi^{2}}\sum\limits_{n=0}^{\infty}\iota_{n}\int\limits_{\omega_{n0}}^{\infty}\frac{{\rm
d}\kappa}{{\rm
e}^{2{\kappa}a}-1}\frac{\kappa^{2}}{\sqrt{\kappa^{2}-\omega_{n0}^{2}}}.\eqno(105)
$$
The integral in (104) can be taken after expanding the factor with
denominator as $\sum\limits_{j=1}^{\infty}{\rm
e}^{-2j{\kappa}a}$. In this way, we obtain the following expressions
for the Casimir energy
$$
\left.\frac{E_{\rm
ren}}{S}\right|_{\tilde\varphi=0}=a\varepsilon^{\infty}_{\rm
ren}+
\frac{|eB|}{\pi^{2}}\sum\limits_{n=0}^{\infty}\iota_{n}\omega_{n0}\sum\limits_{j=1}^{\infty}\frac{1}{j}K_{1}(2j\omega_{n0}a)
-\frac{|eB|}{2\pi}\sum\limits_{n=0}^{\infty}\iota_{n}\omega_{n0} \eqno(106)
$$
and the Casimir pressure
$$
\left.F\right|_{\tilde\varphi=0}=-\varepsilon^{\infty}_{\rm
ren}+
\frac{2|eB|}{\pi^{2}}\sum\limits_{n=0}^{\infty}\iota_{n}\omega_{n0}^{2}\sum\limits_{j=1}^{\infty}\left[K_{0}(2j\omega_{n0}a)+
\frac{1}{2j\omega_{n0}a}K_{1}(2j\omega_{n0}a)\right], \eqno(107)
$$
where $K_{\nu}(s)$ is the Macdonald function of order $\nu$.

It should be noted that the periodic boundary condition,
$$
\chi|_{z=-a/2}=\chi|_{z=a/2}, \quad  \tilde\chi|_{z=-a/2}=\tilde\chi|_{z=a/2}, \eqno(108)
$$
ensures the self-adjointness of the Dirac hamiltonian operator, but current $J^{z}_{qnlj}(\mathbf{r})=\psi_{qnl}^{(j)\dag}(\mathbf{r}){\alpha}^{z}\psi_{qnl}^{(j)}(\mathbf{r})$ $(j=0,1,2)$ does not
vanish at the boundary: instead, the influx of the quantized matter
through a one boundary plane equals the outflux of the quantized matter
through another boundary plane, 
$$
J^{z}_{qnlj}(\mathbf{r})|_{z=-{a/2}}=J^{z}_{qnlj}(\mathbf{r})|_{z={a/2}}. \eqno(109)
$$
The spectrum of the wave number
vector which is orthogonal to the boundary is
$k_{l}=\frac{2\pi}{a}l \,\, (l=0,\pm1,\pm2,...)$, and
the Casimir pressure is
$$
(F)_{\rm periodic}=-\varepsilon^{\infty}_{\rm ren}+
\frac{2|eB|}{\pi^{2}}\sum\limits_{n=0}^{\infty}\iota_{n}\int\limits_{\omega_{n0}}^{\infty}\frac{{\rm
d}\kappa}{{\rm
e}^{{\kappa}a}-1}\frac{\kappa^{2}}{\sqrt{\kappa^{2}-\omega_{n0}^{2}}},\eqno(110)
$$
or, in the alternative representation,
$$
(F)_{\rm periodic}=-\varepsilon^{\infty}_{\rm ren}+
\frac{2|eB|}{\pi^{2}}\sum\limits_{n=0}^{\infty}\iota_{n}\omega_{n0}^{2}\sum\limits_{j=1}^{\infty}\left[K_{0}(j\omega_{n0}a)+
\frac{1}{j\omega_{n0}a}K_{1}(j\omega_{n0}a)\right].\eqno(111)
$$

It is instructive to consider also the case of $\varphi=\pi/2$, when relations (93)-(97) are reduced to
\newpage
$$
\left.\Upsilon(\kappa)\right|_{\varphi=\pi/2}
=-\biggl\{\biggl[\left(2\kappa a-1\right)\bigg(1-
\frac{2\kappa^{2}\sin^{2}\tilde\varphi}{\kappa^{2}-\omega_{n0}^{2}\cos^{2}\tilde\varphi}\biggr)
-\frac{\kappa^{2}\omega_{n0}^{2}\sin^{2}2\tilde\varphi}{\left(\kappa^{2}
-\omega_{n0}^{2}\cos^{2}\tilde\varphi\right)^{2}}\biggr]{\rm e}^{6{\kappa}a}
$$
$$
-\biggl[4\kappa a-3+
\frac{2\kappa^{2}\left(\kappa^{2}-\omega_{n0}^{2}\right)\sin^{2}2\tilde\varphi}{\left(\kappa^{2}
-\omega_{n0}^{2}\cos^{2}\tilde\varphi\right)^{2}}\biggr]{\rm e}^{4{\kappa}a}
$$
$$
+\biggl[\left(2\kappa a-3\right)\bigg(1-
\frac{2\kappa^{2}\sin^{2}\tilde\varphi}{\kappa^{2}-\omega_{n0}^{2}\cos^{2}\tilde\varphi}\biggr)
+\frac{\kappa^{2}\omega_{n0}^{2}\sin^{2}2\tilde\varphi}{\left(\kappa^{2}
-\omega_{n0}^{2}\cos^{2}\tilde\varphi\right)^{2}}\biggr]{\rm e}^{2{\kappa}a}+1\biggr\}
$$
$$
\times\biggl[\left({\rm e}^{2{\kappa}a}-1\right)^{2}+
\frac{4\kappa^{2}\sin^{2}\tilde\varphi}{\kappa^{2}-\omega_{n0}^{2}\cos^{2}\tilde\varphi}{\rm e}^{2{\kappa}a}\biggr]^{-2}. \eqno(112)
$$
This case interpolates between the case of spectrum $k_{l}=\frac{\pi}{a}l \,\, (l=0,1,2,...)$, 
see (99) and (104)-(107), and the case of 
spectrum $k_{l}=\frac{\pi}{a}(l+\frac{1}{2})$ $(l=0,1,2,...)$ with
$$
\left.\Upsilon(\kappa)\right|_{\varphi=-\tilde\varphi=\pi/2}
=\frac{\left(2\kappa a-1\right)
{\rm e}^{2{\kappa}a}-1}
{\left({\rm e}^{2{\kappa}a}+1\right)^{2}} \eqno(113)
$$
and
$$
\left.F\right|_{\varphi=-\tilde\varphi=\pi/2}=-\varepsilon^{\infty}_{\rm
ren}-
\frac{2|eB|}{\pi^{2}}\sum\limits_{n=0}^{\infty}\iota_{n}\int\limits_{\omega_{n0}}^{\infty}\frac{{\rm
d}\kappa}{{\rm
e}^{2{\kappa}a}+1}\frac{\kappa^{2}}{\sqrt{\kappa^{2}-\omega_{n0}^{2}}}, \eqno(114)
$$
or, alternatively,
$$
\left.F\right|_{\varphi=-\tilde\varphi=\pi/2}=-\varepsilon^{\infty}_{\rm
ren}
$$
$$
-\frac{2|eB|}{\pi^{2}}\sum\limits_{n=0}^{\infty}\iota_{n}\omega_{n0}^{2}\sum\limits_{j=1}^{\infty}
(-1)^{j-1}\left[K_{0}(2j\omega_{n0}a)+
\frac{1}{2j\omega_{n0}a}K_{1}(2j\omega_{n0}a)\right]. \eqno(115)
$$

Concluding this section, let us note that in the case of spectrum $k_{l}=\frac{2\pi}{a}(l+\frac{1}{2})$ $(l=0,\pm1,\pm2,...)$, corresponding to 
the antiperiodic boundary condition,
$$
\chi|_{z=-a/2}=-\chi|_{z=a/2}, \quad  \tilde\chi|_{z=-a/2}=-\tilde\chi|_{z=a/2}, \eqno(116)
$$
one  obtains \cite{Si1}
$$
(F)_{\rm antiperiodic}=-\varepsilon^{\infty}_{\rm ren}-
\frac{2|eB|}{\pi^{2}}\sum\limits_{n=0}^{\infty}\iota_{n}\int\limits_{\omega_{n0}}^{\infty}\frac{{\rm
d}\kappa}{{\rm
e}^{{\kappa}a}+1}\frac{\kappa^{2}}{\sqrt{\kappa^{2}-\omega_{n0}^{2}}},\eqno(117)
$$
or, alternatively,
$$
(F)_{\rm antiperiodic}=-\varepsilon^{\infty}_{\rm ren}
$$
$$
-\frac{2|eB|}{\pi^{2}}\sum\limits_{n=0}^{\infty}\iota_{n}\omega_{n0}^{2}\sum\limits_{j=1}^{\infty}(-1)^{j-1}
\left[K_{0}(j\omega_{n0}a)+
\frac{1}{j\omega_{n0}a}K_{1}(j\omega_{n0}a)\right]. \eqno(118)
$$

\section{Asymptotics at small and large separations of plates}

The expression for the Casimir pressure, see (92), can be presented as 
$$
F=-\varepsilon^{\infty}_{\rm ren} + \Delta_{\varphi,\tilde\varphi}(a),  \eqno(119)
$$
where the first term is equal to minus the vacuum energy density which is induced by the magnetic field in  
unbounded space, see (49), whereas the second term which is given by the sum over $n$ and the integral over $\kappa$ in (92) 
depends on the distance between bounding plates and on a choice of boundary conditions at the plates. 

In the case of a weak magnetic field, $|B|\ll{m^{2}|e|^{-1}}$, substituting the sum by integral 
$\int\limits_{0}^{\infty}{\rm d}n$ and changing the integration variable, we get
$$
\Delta_{\varphi,\tilde\varphi}(a)=-
\frac{1}{\pi^{2}}\int\limits_{m}^{\infty}{\rm
d}\kappa(\kappa^{2}-m^{2})^{3/2}\int\limits_{0}^{1}{\rm
d}v\sqrt{1-v}\tilde\Upsilon(\kappa,v),\quad
|eB|\ll{m^{2}},\eqno(120)
$$
where $\tilde\Upsilon(\kappa,v)$ is obtained from $\Upsilon(\kappa)$ (93) by substitution
$\mu_n(\varphi,\tilde\varphi)\rightarrow\tilde\mu_{v,\kappa^{2}}(\varphi,\tilde\varphi)$ with
$$
\tilde\mu_{v,\kappa^{2}}(\varphi,\tilde\varphi)=v(\kappa^{2}-m^{2})\cos^{2}\tilde\varphi
+m^{2}\sin(\varphi+\tilde\varphi)\sin(\varphi-\tilde\varphi). \eqno(121)
$$
In the limit of small distances between the plates, $ma\ll1$, (120) becomes independent of the $\varphi$-parameter:
$$
\Delta_{\varphi,\tilde\varphi}(a)=
\frac{1}{4a^{4}}\biggl\{\frac{\pi^{2}}{30} - \int\limits_{0}^{1}{\rm d}v \, \rho_{\tilde\varphi}(v)
\biggl(1 - \frac{|\rho_{\tilde\varphi}(v)|}{\pi}\biggr)\biggl[\frac{3}{2}\sqrt{1-v} \, \rho_{\tilde\varphi}(v)
\biggl(1 - \frac{|\rho_{\tilde\varphi}(v)|}{\pi}\biggr)
$$
$$
+ \frac{v\sin2\tilde\varphi}{1-v\cos^{2}\tilde\varphi}
\biggl(\frac{1}{2} - \frac{|\rho_{\tilde\varphi}(v)|}{\pi}\biggr)\biggr]\biggr\}, \quad \sqrt{|eB|}a \ll ma \ll 1, \eqno(122)
$$
where
$$
\rho_{\tilde\varphi}(v)= \arcsin \biggl(\frac{\sin\tilde\varphi}{\sqrt{1-v\cos^{2}\tilde\varphi}}\biggr). \eqno(123)
$$
Thus, $\Delta_{\varphi,\tilde\varphi}(a)$ in this case is power-dependent on the distance between the plates as $a^{-4}$ with 
the dimensionless constant of proportionality, either positive or negative, depending on the value of the 
$\tilde\varphi$-parameter. In particular, we get
$$
\Delta_{\varphi,0}(a)=\frac{\pi^{2}}{120}\frac{1}{a^{4}}, \quad \sqrt{|eB|}a \ll ma \ll 1 \eqno(124)
$$
and
$$
\Delta_{\varphi,{-\pi}/{2}}(a)=-\frac{7}{8}\frac{\pi^{2}}{120}\frac{1}{a^{4}}, 
\quad \sqrt{|eB|}a \ll ma \ll 1. \eqno(125)
$$
In the limit of large distances between the plates, $ma\gg1$, $\Delta_{\varphi,\tilde\varphi}(a)$ (120) takes form
$$
\Delta_{\varphi,\tilde\varphi}(a)=
\frac{2}{\pi^{2}}\int\limits_{m}^{\infty}{\rm
d}\kappa\kappa(\kappa^{2}-m^{2})^{3/2}{\rm e}^{-2{\kappa}a}\int\limits_{0}^{1}{\rm
d}v\sqrt{1-v}
$$
$$
\times\biggl\{a\frac{\kappa^2\cos2\tilde\varphi-\tilde\mu_{v,\kappa^2}(\varphi,\tilde\varphi)}
{{\kappa}^{2}-2{\kappa}m\cos\varphi\sin\tilde\varphi-\tilde\mu_{v,\kappa^2}(\varphi,\tilde\varphi)}
$$
$$
-\frac{(2{\kappa}\sin\tilde\varphi-m\cos\varphi)\tilde\mu_{v,\kappa^2}(\varphi,\tilde\varphi)
-{\kappa}^{2}m\cos\varphi\cos2\tilde\varphi}
{[{\kappa}^{2}-2{\kappa}m\cos\varphi\sin\tilde\varphi-\tilde\mu_{v,\kappa^2}(\varphi,\tilde\varphi)]^{2}}\sin\tilde\varphi\biggr\},
$$
$$
|eB|\ll{m^{2}}, \quad  ma\gg1. \eqno(126)
$$
Clearly, (126) is suppressed as ${\rm exp}(-2ma)$. In particular, we get
$$
\Delta_{\varphi,0}(a)=\frac{1}{2\pi^{3/2}}\frac{m^{5/2}}{a^{3/2}}{\rm
e}^{-2ma}\left[1+O\left(\frac{1}{ma}\right)\right], \, |eB|\ll{m^{2}}, \, ma\gg1  \eqno(127)
$$
and
$$
\Delta_{\varphi,{-\pi}/{2}}(a)=
\left\{\begin{array}{l}-\frac{3}{16\pi^{3/2}}\frac{m^{3/2}}{a^{5/2}}{\rm
e}^{-2ma}[1+O(\frac{1}{ma})],\quad \varphi=0
\\ [6 mm]
-\frac{\tan^{2}(\varphi/2)}{2\pi^{3/2}}\frac{m^{5/2}}{a^{3/2}}{\rm
e}^{-2ma}[1+O(\frac{1}{ma})],\quad \varphi \neq 0
\end{array}
\right\}, 
$$
$$
|eB|\ll{m^{2}}, \quad  ma\gg1.  \eqno(128)
$$

In the case of a strong magnetic field, $|B|\gg{m}^{2}|e|^{-1}$, one has
$$
\Delta_{\varphi,\tilde\varphi}(a)=-
\frac{|eB|}{\pi^{2}}\left[\int\limits_{m}^{\infty}{\rm
d}\kappa\sqrt{\kappa^{2}-m^{2}}\Upsilon(\kappa)|_{n=0}\right.
$$
$$
\left. +2\sum\limits_{n=1}^{\infty}\int\limits_{\sqrt{2n|eB|}}^{\infty}{\rm d}\kappa\sqrt{\kappa^{2}-2n|eB|}
\Upsilon(\kappa)|_{m=0}\right], \quad |eB|\gg{m}^{2}. \eqno(129)
$$
In the limit of extremely small distances between the plates, 

\noindent $ma\ll\sqrt{|eB|}a\ll1$, the analysis is 
similar to that of the limit of 

\noindent $\sqrt{|eB|}a \ll ma \ll 1$, yielding the same results as (122)-(125). Otherwise, in the 
limit of $\sqrt{|eB|}a\gg1$, only the first term in square brackets on the right-hand side of (129) matters. In the limit 
of small distances between the plates this term becomes $\varphi$-independent, yielding
$$
\Delta_{\varphi,\tilde\varphi}(a)=
\frac{|eB|}{4a^{2}}\left[\frac{1}{6}-\frac{|\tilde\varphi|}{\pi}\left(1-\frac{|\tilde\varphi|}{\pi}\right)\right], 
\quad \sqrt{|eB|}a\gg1, \, ma\ll1. \eqno(130)
$$
In particular, we get
$$
\Delta_{\varphi,0}(a)=\frac{|eB|}{24a^{2}}, \quad
\sqrt{|eB|}a \gg 1, \, ma \ll 1, \eqno(131)
$$
$$
\Delta_{\varphi,{\pm\pi}/{4}}(a)=- \frac{|eB|}{192a^{2}},\quad
\sqrt{|eB|}a\gg1, \, ma\ll1 \eqno(132)
$$
and
$$
\Delta_{\varphi,{-\pi}/{2}}(a)=- \frac{|eB|}{48a^{2}},\quad
\sqrt{|eB|}a\gg1, \, ma\ll1. \eqno(133)
$$
In the limit of large distances between the plates, the first term in square brackets on the right-hand side of (129) yields
\newpage
$$
\Delta_{\varphi,\tilde\varphi}(a)=
\frac{2|eB|}{\pi^{2}}\int\limits_{m}^{\infty}{\rm
d}\kappa\kappa(\kappa^{2}-m^{2})^{1/2}{\rm e}^{-2{\kappa}a}
$$
$$
\times\biggl\{a\frac{\kappa^2\cos2\tilde\varphi-m^{2}\sin(\varphi+\tilde\varphi)\sin(\varphi-\tilde\varphi)}
{{\kappa}^{2}-2{\kappa}m\cos\varphi\sin\tilde\varphi-m^{2}\sin(\varphi+\tilde\varphi)\sin(\varphi-\tilde\varphi)}
$$
$$
+\frac{{\kappa}^{2}m\cos\varphi\cos2\tilde\varphi
-(2{\kappa}\sin\tilde\varphi-m\cos\varphi)m^{2}\sin(\varphi+\tilde\varphi)\sin(\varphi-\tilde\varphi)}
{[{\kappa}^{2}-2{\kappa}m\cos\varphi\sin\tilde\varphi-m^{2}\sin(\varphi+\tilde\varphi)\sin(\varphi-\tilde\varphi)]^{2}}
\sin\tilde\varphi\biggr\},
$$
$$
\sqrt{|eB|}a\gg{ma}\gg1, \eqno(134)
$$
which is obviously suppressed as ${\rm exp}(-2ma)$. In particular, we get
$$
\Delta_{\varphi,0}(a)=
\frac{|eB|}{2\pi^{3/2}}\frac{m^{3/2}}{a^{1/2}}{\rm
e}^{-2ma}\left[1+O\left(\frac{1}{ma}\right)\right],\quad
\sqrt{|eB|}a\gg{ma}\gg1 \eqno(135)
$$
and
$$
\Delta_{\varphi,{-\pi}/{2}}(a)=
\left\{\begin{array}{l}-\frac{|eB|}{16\pi^{3/2}}\frac{m^{1/2}}{a^{3/2}}{\rm
e}^{-2ma}[1+O(\frac{1}{ma})],\quad \varphi=0
\\ [6 mm]
-\frac{|eB|\tan^{2}(\varphi/2)}{2\pi^{3/2}}\frac{m^{3/2}}{a^{1/2}}{\rm
e}^{-2ma}[1+O(\frac{1}{ma})],\quad \varphi \neq 0 \end{array}
\right\},
$$
$$
\sqrt{|eB|}a\gg{ma}\gg1. \eqno(136)
$$

It is appropriate in this section to consider also the limiting case of $m\rightarrow0$. In view of the asymptotical behaviour 
of the boundary-independent piece of the Casimir pressure,
$$
-\varepsilon^{\infty}_{\rm
ren}=\frac{e^{2}B^{2}}{24\pi^{2}}\ln\frac{2|eB|}{m^{2}}, \quad m^{2}\ll|eB| \eqno(137)
$$
and
$$
-\varepsilon^{\infty}_{\rm
{ren}}=\frac{1}{360\pi^{2}}\frac{e^{4}B^{4}}{m^{4}}, \quad m^{2}\gg|eB|, \eqno(138)
$$
namely asymptotics (122) is relevant for this case, and the pressure from the vacuum of a confined massless 
spinor matter field is given by expression 
\newpage
$$
\left.F\right|_{m=0, \, B=0}=
\frac{1}{8a^{4}}\biggl\{\frac{\pi^{2}}{30} - \int\limits_{0}^{1}{\rm
d}v \, \rho_{\tilde\varphi}(v)\biggl(1 - \frac{|\rho_{\tilde\varphi}(v)|}{\pi}\biggr)
$$
$$
\times\biggl[\frac{3}{2}\sqrt{1-v} \, \rho_{\tilde\varphi}(v)\biggl(1 - \frac{|\rho_{\tilde\varphi}(v)|}{\pi}\biggr) 
+ \frac{v\sin2\tilde\varphi}{1-v\cos^{2}\tilde\varphi}
\biggl(\frac{1}{2} - \frac{|\rho_{\tilde\varphi}(v)|}{\pi}\biggr)\biggr]\biggr\}, \eqno(139)
$$
which is bounded from above and below by values
$$
\left.F\right|_{m=0, \, B=0, \, \tilde\varphi= 0}=\frac{\pi^{2}}{240}\frac{1}{a^{4}} \eqno(140)
$$
and
$$
\left.F\right|_{m=0, \, B=0, \, \tilde\varphi={-\pi}/{2}}=-\frac{7}{8}\frac{\pi^{2}}{240}\frac{1}{a^{4}}, \eqno(141)
$$
respectively; here, an additional factor of 1/2 has appeared due to diminishment of the number of degrees 
of freedom (a massless spinor can be either left or right).

\section{Summary and discussion}

In the present paper, we consider an impact of a background (classical) magnetic field on the
vacuum of a quantized charged spinor matter field which is confined
to a bounded region of space; the sources of the magnetic field are outside of the bounded region, and the magnetic 
field strength lines are assumed to be orthogonal to a boundary. The confinement of the matter field 
(i.e. absence of the matter flux across the boundary) is ensured by boundary condition (34) which is 
compatible with the self-adjointness of the Dirac hamiltonan operator and which generalizes the well-known 
MIT bag boundary condition to the most extent; the parameters of this general boundary condition can 
be interpreted as the self-adjoint extension parameters. In the case which is relevant to the geometry of 
the Casimir effect (i.e. the spatial region bounded by two parallel planes separated by distance $a$) and the uniform magnetic 
field orthogonal to the planes, the eight-parameter boundary condition is reduced to the four-parameter one, 
see (37). With the use of the latter condition we obtain the equation determining the spectrum of the 
wave number vector along the magnetic field, see (69)-(71). Following the analysis of this equation, we 
finally arrive at the two-parameter boundary condition, see (81), as the most general extension of the 
MIT bag boundary condition in the context of the Casimir effect. The spectrum of the wave number vector along 
the magnetic field in this case depends on the number of the Landau level and on the sign of the 
one-particle energy, see (80). The Casimir pressure is shown to take form of (92), where  
$\varepsilon^{\infty}_{\rm {ren}}$ is given by (49) and $\Upsilon(\kappa)$ is given by (93)-(97). The result 
for the case of the MIT bag boundary condition is obtained from (92) at $\varphi=0$, $\tilde\varphi={-\pi}/{2}$. 
It should be noted that the periodic and antiperiodic boundary conditions, see (108) and (116), do not ensure 
the confinement of the matter field: instead, they correspond to the equality between the matter influx through a one 
boundary plane and the matter outflux through another boundary plane, see (109). Nevertheless, the Casimir pressure in these two cases is obtainable by 
substitution $a\rightarrow{a/2}$ in (92) at $\tilde\varphi=0$ (periodic boundary condition) and at 
$\varphi=-\tilde\varphi=\pi/2$ (antiperiodic boundary condition), see (110)-(111) and (117)-(118), respectively.

The Casimir effect is usually validated in experiments with 
nearly parallel plates separated by a distance of order $10^{-8}-10^{-5} \, \rm m$, see, e.g., \cite{Bor}. 
The Compton wavelength of the lightest charged particle, electron, is $m^{-1} \sim 10^{-12}\,\rm m$, thus ${ma}\gg1$ 
and, as has been shown in the preceding section, all the dependence of the Casimir pressure on the distance between the plates 
and a choice of boundary conditions at the plates is suppressed by factor ${\rm exp}(-2ma)$, see (126)-(128) and (134)-(136). 
Hence, the pressure from the electron-positron vacuum onto the plates separated by distance $a>10^{-10}\,\rm m$ 
is well approximated by $F \approx -\varepsilon^{\infty}_{\rm {ren}}$, where $\varepsilon^{\infty}_{\rm {ren}}$ (49) is 
negative, i.e. the pressure is positive and the plates are repelled. Some possibilities to detect this new-type Casimir 
effect were pointed out in \cite{Si1}.

Let us also discuss an application of our results to hadron physics. Since the hadron size (confinement radius) is 
$a \sim 10^{-15}\,\rm m$ and the Compton wavelength of the lightest quark is $m^{-1} \sim 10^{-13}\,\rm m$, it looks likely that 
asymptotical regime ${ma}\ll1$ might be relevant. Although the geometry of compact bounded space (for instance, of a sphere) 
is more appropriate to this case, we would like just to emphasize here that the behaviour of the pressure from the vacuum of 
the confined quark-antiquark matter differs drastically for weak and strong magnetic fields; this has been shown in the present paper for 
the geometry of noncompact space bounded by two parallel planes and the uniform magnetic field orthogonal to the planes. If 
$-\varepsilon^{\infty}_{\rm {ren}} \ll |\Delta_{\varphi,\tilde\varphi}(a)|$, then the pressure is 
$F \approx \Delta_{\varphi,\tilde\varphi}(a)$ which can be either positive or negative, depending on boundary conditions and 
independent of the magnetic field strength, see (122)-(125); this case
\footnote{Note that the pressure equals twice the pressure from the vacuum of the confined massless neutral spinor matter, 
see (139).} 
is relevant for $|B| < 10^{13} \, \rm Gauss$. If $-\varepsilon^{\infty}_{\rm {ren}} \gg |\Delta_{\varphi,\tilde\varphi}(a)|$, 
then the pressure is 
$F \approx -\varepsilon^{\infty}_{\rm {ren}}$ which is positive and independent of boundary 
conditions, see (137); this case is relevant for $|B| > 10^{19} \, \rm Gauss$. The pressure for the intermediate magnetic fields, 
$10^{13} \, \rm Gauss < |B| < 10^{19} \, \rm Gauss$, depends both on boundary conditions and on the 
magnetic field strength. The magnetic fields of strength up to $10^{17} - 10^{18} \, \rm Gauss$ may exist in some compact 
astrophysical objects (magnetars) \cite{Mer}, while even much stronger magnetic fields are supposed to 
have existed in early universe \cite{Gra}.

\section*{Acknowledgments}

We acknowledge the support from the National Academy of Sciences
of Ukraine (project No.0112U000054). The work of Yu.~A.~S. was supported by the
Program of Fundamental Research of the Department of Physics and
Astronomy of the National Academy of Sciences of Ukraine (project
No.0112U000056) and by the ICTP -- SEENET-MTP grant PRJ-09
``Strings and Cosmology''.

\section*{Appendix A. Solution to the Dirac equation in a background uniform magnetic field}

A solution to the Dirac equation in the background of a static uniform magnetic field is
well-described in the literature, see, e.g., \cite{Akhie}. Taking $eB>0$ for definiteness,
the solution with positive energy is
$$
\left.\psi_{qnk}(\mathbf{r})\right|_{E_{nk}=\omega_{nk}}=\frac{{\rm e}^{{\rm i}qx}{\rm e}^{{\rm i}kz}}{2\pi\sqrt{2\omega_{nk}(\omega_{nk}+m)}}\left[C_{1}\begin{pmatrix} (\omega_{nk}+m)Y_{n}^{q}(y) \\ 0 \\ kY_{n}^{q}(y) \\ \sqrt{2neB} Y_{n-1}^{q}(y) \end{pmatrix}\right.$$
$$\left.+C_{2}\begin{pmatrix} 0 \\ (\omega_{nk}+m)Y_{n-1}^{q}(y) \\
\sqrt{2neB} Y_{n}^{q}(y) \\ -kY_{n-1}^{q}(y)
\end{pmatrix}\right],\quad n\geq1 \eqno(A.1)
$$
and
$$
\left.\psi^{(0)}_{q0k}(\mathbf{r})\right|_{E_{0k}=\omega_{0k}}=\frac{{\rm e}^{{\rm i}qx}{\rm e}^{{\rm i}kz}}{2\pi\sqrt{2\omega_{0k}(\omega_{0k}+m)}}
C_{0}Y_{0}^{q}(y)\begin{pmatrix} \omega_{0k}+m \\ 0 \\ k \\ 0 \end{pmatrix}, \eqno(A.2)
$$
while the solution with negative energy is 
$$
\left.\psi_{qnk}(\mathbf{r})\right|_{E_{nk}=-\omega_{nk}}=\frac{{\rm e}^{-{\rm i}qx}{\rm e}^{-{\rm i}kz}}{2\pi\sqrt{2\omega_{nk}(\omega_{nk}+m)}}
\left[\tilde{C}_{1}\begin{pmatrix} kY_{n}^{-q}(y)
\\-\sqrt{2neB}Y_{n-1}^{-q}(y) \\ (\omega_{nk}+m)Y_{n}^{-q}(y) \\ 0
\end{pmatrix}\right.
$$
$$\left. + \tilde{C}_{2}\begin{pmatrix}-\sqrt{2neB}Y_{n}^{-q}(y)\\
-kY_{n-1}^{-q}(y) \\ 0 \\(\omega_{nk}+m)Y_{n-1}^{-q}(y)
\end{pmatrix}\right],\quad n\geq1\eqno(A.3)
$$
and
$$
\left.\psi^{(0)}_{q0k}(\mathbf{r})\right|_{E_{0k}=-\omega_{0k}}=
\frac{{\rm e}^{-{\rm i}qx}{\rm e}^{-{\rm i}kz}}{2\pi\sqrt{2\omega_{0k}(\omega_{0k}+m)}}\tilde{C}_{0}Y_{0}^{-q}(y)
\begin{pmatrix} k \\ 0 \\ \omega_{0k}+m \\ 0 \end{pmatrix}; \eqno(A.4)
$$
here $-\infty<q<\infty$ and
$$
Y_{n}^{q}(y)=\sqrt{\frac{(eB)^{1/2}}{2^{n}n!\pi^{1/2}}}\exp{\left[-\frac{eB}{2}\left(y+\frac{q}{eB}\right)^{2}\right]
H_{n}\left[\sqrt{eB}\left(y+\frac{q}{eB}\right)\right]},\eqno(A.5)
$$ 
$H_n(v)=(-1)^{n}{\rm e}^{v^{2}}\frac{{\rm d}^{n}}{{\rm d}v^{n}}{\rm e}^{-v^{2}}$ is the Hermite polynomial. The case of $eB<0$ is obtained by charge
conjugation, i.e. changing $eB\rightarrow{-eB}$ and multiplying
the complex conjugates  of the previous solutions by ${\rm i}
\beta {\alpha}^2$ (the energy sign is changed to the opposite).

Solutions with different signs of energy are orthogonal:
$$
\int{\rm d}^{3}r\,\psi^{(j)\dag}_{-qn-k}(\mathbf{r})|_{E_{nk}=-\omega_{nk}}
\psi_{q'n'k'}^{(j')}(\mathbf{r})|_{E_{n'k'}=\omega_{n'k'}}=0. \eqno(A.6)
$$
In the case of $n\geq1$, two linearly independent solutions with, say, positive energy,
$\left.\psi_{qnk}^{(1)}(\mathbf{r})\right|_{E_{nk}=\omega_{nk}}$ and $\left.\psi_{qnk}^{(2)}(\mathbf{r})\right|_{E_{nk}=\omega_{nk}}$,
are orthogonal, if the appropriate coefficients, $C_{j}^{(1)}$ and
$C_{j}^{(2)}(j=1,2)$, obey condition
$$
\sum\limits_{j=1,2}C_{j}^{(1)*}C_{j}^{(2)}=0.\eqno(A.7)
$$
We impose further condition 
$$
\sum\limits_{j=1,2}|C_{j}^{(j')}|^{2}=|C_{0}|^{2}=1,\quad
j'=1,2. \eqno(A.8)
$$
The same conditions are demanded for coefficients $\tilde{C}_{0}$ and $\tilde{C}_{j}^{(j')} \quad (j,j'=1,2)$ corresponding to the case of 
the negative-energy solutions. Then the wave functions satisfy the requirements of
orthonormality
$$\int{\rm d}^{3}r\,\psi^{(j)\dag}_{qnk}(\mathbf{r})\psi_{q'n'k'}^{(j')}(\mathbf{r})=\delta_{jj'}\delta_{nn'}\delta(q-q')\delta(k-k'), \quad
j,j'=0,1,2 \eqno(A.9)
$$
and completeness
$$
\sum\limits_{{\rm sgn}(E_{nk})}\int\limits_{-\infty}^{\infty}{\rm{d}}q\int\limits_{-\infty}^{\infty}{\rm{d}}k\left[\psi^{(0)}_{q0k}(\mathbf{r})
\psi^{(0)\dag}_{q0k}(\mathbf{r'})+\sum\limits_{n=1}^{\infty}\sum\limits_{j=1,2}\psi^{(j)}_{qnk}(\mathbf{r})
\psi^{(j)\dag}_{qnk}(\mathbf{r'})\right]=I\delta(\mathbf{r}-\mathbf{r'}).\eqno(A.10)
$$
With the use of relation
$$\int\limits_{-\infty}^{\infty}{\rm{d}}q\,\left[Y^{q}_{n}(y)\right]^2=|eB|,\eqno(A.11)
$$
expression (48) is readily obtained.

The solution corresponding to a plane wave propagating along the $z$-axis in
the opposite direction is written in the case of $eB>0$ in the following form:
\newpage
$$
\left.\psi_{qn-k}(\mathbf{r})\right|_{E_{nk}=\omega_{nk}}=\frac{{\rm e}^{{\rm i}qx}{\rm e}^{-{\rm i}kz}}{2\pi\sqrt{2\omega_{nk}(\omega_{nk}+m)}}\left[\tilde{C}_{1}\begin{pmatrix} (\omega_{nk}+m)Y_{n}^{q}(y) \\ 0 \\ -kY_{n}^{q}(y) \\ \sqrt{2neB}Y_{n-1}^{q}(y) \end{pmatrix}\right.$$
$$\left.+\tilde{C}_{2}\begin{pmatrix} 0 \\ (\omega_{nk}+m)Y_{n-1}^{q}(y) \\
\sqrt{2neB}Y_{n}^{q}(y) \\ kY_{n-1}^{q}(y)
\end{pmatrix}\right], \quad n\geq1  \eqno(A.12)
$$
and
$$
\left.\psi^{(0)}_{q0-k}(\mathbf{r})\right|_{E_{0k}=\omega_{0k}}=
\frac{{\rm e}^{{\rm i}qx}{\rm e}^{-{\rm i}kz}}{2\pi\sqrt{2\omega_{0k}(\omega_{0k}+m)}}\tilde{C}_{0}Y_{0}^{q}(y)
\begin{pmatrix} \omega_{0k}+m \\ 0 \\ -k \\ 0 \end{pmatrix}, \eqno(A.13)
$$
or, for the opposite sign of energy, 
$$
\left.\psi_{qn-k}(\mathbf{r})\right|_{E_{nk}=-\omega_{nk}}=
\frac{{\rm e}^{-{\rm i}qx}{\rm e}^{{\rm i}kz}}{2\pi\sqrt{2\omega_{nk}(\omega_{nk}+m)}}
\left[{C}_{1}\begin{pmatrix} -kY_{n}^{-q}(y)
\\-\sqrt{2neB}Y_{n-1}^{-q}(y) \\ (\omega_{nk}+m)Y_{n}^{-q}(y) \\ 0
\end{pmatrix}\right.
$$
$$\left. + {C}_{2}\begin{pmatrix}-\sqrt{2neB}Y_{n}^{-q}(y)\\
kY_{n-1}^{-q}(y) \\ 0 \\(\omega_{nk}+m)Y_{n-1}^{-q}(y)
\end{pmatrix}\right],\quad n\geq1 \eqno(A.14)
$$
and
$$
\left.\psi^{(0)}_{q0-k}(\mathbf{r})\right|_{E_{0k}=-\omega_{0k}}=
\frac{{\rm e}^{-{\rm i}qx}{\rm e}^{{\rm i}kz}}{2\pi\sqrt{2\omega_{0k}(\omega_{0k}+m)}}{C}_{0}Y_{0}^{-q}(y)
\begin{pmatrix} -k \\ 0 \\ \omega_{0k}+m \\ 0 \end{pmatrix}. \eqno(A.15)
$$

\section*{Appendix B. Abel-Plana summation formula}

Condition (80) is rewritten as
$$
P_{+}(k_{l})=0, \quad  E_{nl}>0, \eqno(B.1)
$$
or
$$
P_{-}(k_{l})=0, \quad  E_{nl}<0, \eqno(B.2)
$$
where
$$
P_{\pm}(u)=\cos(ua)+\frac{\pm\omega_{nu}\cos\tilde{\varphi}-m\cos\varphi}{u\sin\tilde{\varphi}}\sin(ua) \eqno(B.3)
$$
and, see (47),
$$
\omega_{nu}=\sqrt{u^{2}+2n|eB|+m^{2}}. \eqno(B.4)
$$
We assign labels $l=0,1,2,...$, to the consecutively increasing
positive roots of (B.1) or (B.2), $k_{l}>0$; appropriately, labels
$l=-1,-2,...$, are assigned to the consecutively decreasing
negative roots of (B.1) or (B.2), $k_{l}<0$. Then one can write
$$
\sum\limits_{{\rm sgn}(E_{nl})}\sum\limits_{l=0}^{\infty}f(k_{l}^{2})=\frac{1}{2}\sum\limits_{{\rm sgn}(E_{nl})}\sum\limits_{l=-\infty}^{\infty}f(k_{l}^{2})=
\frac{a}{4\pi}\int\limits_{C_{\underline{\overline{\,\,\,\,}}}}{\rm d}u{f(u^{2})}[G_{+}(u)+G_{-}(u)],\eqno(B.5)
$$
where
$$
G_{\pm}(u)=1+\frac{\rm i}{a}\frac{\rm d}{{\rm
d}u}\ln P_{\pm}(u) \eqno(B.6)
$$
and contour $C_{\underline{\overline{\,\,\,\,}}}$ on the complex
$u$-plane consists of two parallel infinite lines going closely
on the lower and upper sides of the real axis, see Fig.1.
\begin{figure}
\includegraphics[width=390pt]{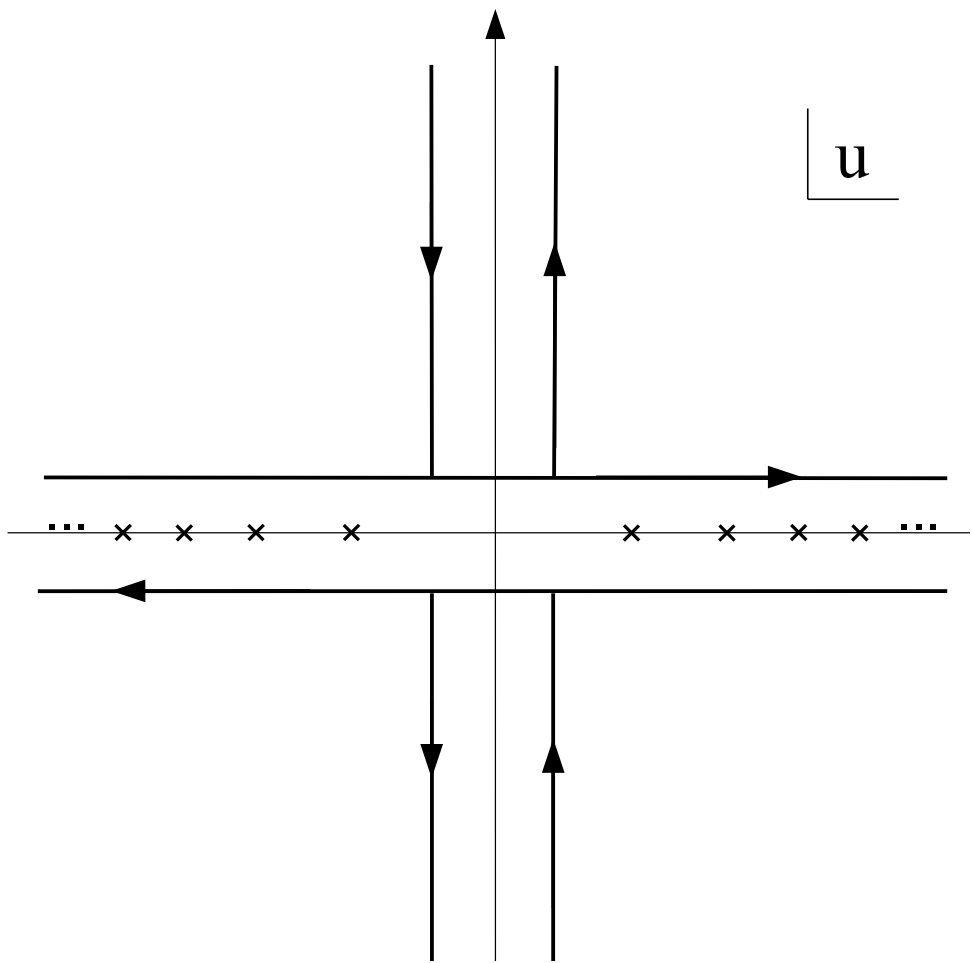}\\
\caption{Contours $C_{\underline{\overline{\,\,\,\,}}}$, $C_{\sqcap}$ and $C_{\sqcup}$ on
the complex $u$-plane; the positions of poles of function $G_{+}(u)+G_{-}(u)$ are
indicated by crosses.}\label{1}
\end{figure}\normalsize
An explicit form of functions $G_{+}(u)$ and $G_{-}(u)$ is
$$
G_{\pm}(u)=\frac{[u+{\rm i}h_{\pm}(u)]{\rm e}^{-{\rm
i}ua} + {\rm i} \sin(ua)\frac{u}{a}\frac{\rm d}{{\rm d}u}[\frac{h_{\pm}(u)}{u}]}{u\cos(ua)+h_{\pm}(u) \sin(ua)}, \eqno(B.7)
$$
where
$$
h_{\pm}(u)=\frac{\pm\omega_{nu}\cos\tilde\varphi - m\cos\varphi}{\sin\tilde\varphi}.\eqno(B.8)
$$
Since the numerators of $G_{+}(u)$ and $G_{-}(u)$ contribute to the integral in (B.5) at values $u=k_l$ only, one may 
change the numerators with the use of relations (B.1) and (B.2), respectively. In this way we define
$$
\tilde{G}_{\pm}(u)=\frac{[u+{\rm i}h_{\pm}(u)]{\rm e}^{-{\rm
i}ua} + {\rm i} g_{\pm}(u)}{u\cos(ua)+h_{\pm}(u) \sin(ua)},\eqno(B.9)
$$
where
$$
g_{\pm}(u)=\frac{\left(\pm\frac{\omega^2_{n0}}{\omega_{nu}}\cos\tilde{\varphi}-m\cos\varphi\right)h_{\pm}(u)}
{a\cos(ua)\sin\tilde\varphi[u^2+h^2_{\pm}(u)]} .\eqno(B.10)
$$
However, additional simple poles at $u=0$ and
at $\cos(ua)=0$ are in $\tilde{G}_{\pm}(u)$ (B.9), as compared to ${G}_{\pm}(u)$ (B.7). Subtracting the contribution of these poles,
we obtain identity
$$
\int\limits_{C_{\underline{\overline{\,\,\,\,}}}}{\rm
d}u{f(u^{2})}[G_+(u)+G_-(u)]=\int\limits_{C_{\underline{\overline{\,\,\,\,}}}}{\rm
d}u{f(u^{2})}[\tilde{G}_+(u)+\tilde{G}_-(u)-\frac{2{\rm i}}{ua}+D_+(u)+D_-(u)], \eqno(B.11)
$$
where
$$
D_{\pm}(u)=\frac{{\rm e}^{-{\rm i}ua}}{a\cos(ua)} d_{\pm}(u), \eqno(B.12)
$$
$$
d_{\pm}(u)=\frac{\pm\frac{\omega^2_{n0}}{\omega_{nu}}\cos\tilde\varphi - m\cos\varphi}{\sin\tilde\varphi[u^2+h^2_{\pm}(u)]}.\eqno(B.13)
$$
Separating explicitly the contribution of the subtracted pole at $u=0$, we consider the remaining integral on the right-hand side
of (B.11) by deforming the parts of contour $C_{\underline{\overline{\,\,\,\,}}}$ into the lower and upper parts of the $u$-plane. 
Note that function $\tilde{G}_+(u)+D_+(u)+\tilde{G}_-(u)+D_-(u)$, in addition to poles at $u=k_l$, has also poles at 
$u=\pm{\rm i}(m\cos\varphi\sin\tilde\varphi + \cos\tilde\varphi\sqrt{2n|eB|+m^{2}\sin^{2}\varphi})$ and 
$u=\pm{\rm i}(m\cos\varphi\sin\tilde\varphi - \cos\tilde\varphi\sqrt{2n|eB|+m^{2}\sin^{2}\varphi})$. Assuming that all 
singularities of $f(u^{2})$ as a function of complex variable $u$ lie on the imaginary axis at some
distances from the origin, we get
$$
\int\limits_{C_{\underline{\overline{\,\,\,\,}}}}{\rm
d}u{f(u^{2})}[G_+(u)+G_-(u)]=\int\limits_{C_{\sqcap}}{\rm d}u{f(u^{2})}[\tilde{G}_+(u)+D_+(u)+\tilde{G}_-(u)+D_-(u)]
$$
$$
+\int\limits_{C_{\sqcup}}{\rm
d}u{f(u^{2})}[\tilde{G}_+(u)+D_+(u)+\tilde{G}_-(u)+D_-(u)]-\frac{4\pi}{a}f(0),\eqno(B.14)
$$
where contours $C_{\sqcap}$ and $C_{\sqcup}$ enclose the lower and upper imaginary semiaxes,
see Fig.1. In view of obvious relation
$$\lim\limits_{k\rightarrow{0+}}(k\pm{{\rm i}\kappa})^{2}=\lim\limits_{k\rightarrow{0+}}(-k\mp{{\rm i}\kappa})^{2}=(\pm{{\rm i}\kappa})^{2}
$$
for real positive $k$ and $\kappa$, we obtain
\newpage
$$
\int\limits_{C_{\underline{\overline{\,\,\,\,}}}}{\rm
d}u{f(u^{2})}[G_+(u)+G_-(u)]={\rm i}\int\limits_{0}^{\infty}{\rm{d}}\kappa\{f[(-{\rm i}\kappa)^{2}]-f[({\rm i}\kappa)^{2}]\}
$$
$$
\times[\tilde G_+(-{\rm i}\kappa)+D_+(-{\rm i}\kappa)+\tilde G_-(-{\rm i}\kappa)+D_-(-{\rm i}\kappa)
$$
$$
-\tilde G_+({\rm i}\kappa)-D_+({\rm i}\kappa)-\tilde G_-({\rm i}\kappa)-D_-({\rm i}\kappa)]-\frac{4\pi}{a}f(0).\eqno(B.15)
$$
Taking account for relations
$$
\tilde G_+(-{\rm i}\kappa)+\tilde G_-(-{\rm i}\kappa)+\tilde G_+({\rm i}\kappa)+\tilde G_-({\rm i}\kappa)=4 \eqno(B.16)
$$
and
$$
D_+(-{\rm i}\kappa)+D_-(-{\rm i}\kappa)+D_+({\rm i}\kappa)+D_-({\rm i}\kappa)
$$
$$
=\frac{2}{a}[d_{+}(-{\rm i}\kappa)+d_{-}(-{\rm i}\kappa)]\equiv\frac{2}{a}[d_{+}({\rm i}\kappa)+d_{-}({\rm i}\kappa)], \eqno(B.17)
$$
we further obtain
$$
\int\limits_{C_{\underline{\overline{\,\,\,\,}}}}{\rm
d}u{f(u^{2})}[G_+(u)+G_-(u)]=2{\rm i}\int\limits_{0}^{\infty}{\rm{d}}\kappa\{f[(-{\rm i}\kappa)^{2}]-f[({\rm i}\kappa)^{2}]\}
$$
$$
\times[\tilde G_+(-{\rm i}\kappa)+D_+(-{\rm i}\kappa)+\tilde G_-(-{\rm i}\kappa)+D_-(-{\rm i}\kappa)]
$$
$$
-2{\rm i}\int\limits_{0}^{\infty}{\rm{d}}{\kappa}f[(-{\rm i}\kappa)^{2})]\left\{2+\frac{1}{a}[d_{+}(-{\rm i}\kappa)
+d_{-}(-{\rm i}\kappa)]\right\}
$$
$$
+2{\rm i}\int\limits_{0}^{\infty}{\rm{d}}{\kappa}f[({\rm i}\kappa)^{2})]\left\{2+\frac{1}{a}[d_{+}({\rm i}\kappa)
+d_{-}({\rm i}\kappa)]\right\}-\frac{4\pi}{a}f(0). \eqno(B.18)
$$
By rotating the paths of integration in the last and before the last
integrals in (B.18) by $90^{\circ}$ in the clockwise and
anticlockwise directions, respectively, we finally get
\newpage
$$
\int\limits_{C_{\underline{\overline{\,\,\,\,}}}}{\rm
d}u{f(u^{2})}[G_+(u)+G_-(u)]=2{\rm i}\int\limits_{0}^{\infty}{\rm{d}}\kappa\{f[(-{\rm i}\kappa)^{2}]-f[({\rm i}\kappa)^{2}]\}
$$
$$
\times[\tilde G_+(-{\rm i}\kappa)+D_+(-{\rm i}\kappa)+\tilde G_-(-{\rm i}\kappa)+D_-(-{\rm i}\kappa)]
$$
$$
+4\int\limits_{0}^{\infty}{\rm{d}}k f(k^{2})\left\{2+\frac{1}{a}[d_{+}(k)
+d_{-}(k)]\right\}-\frac{4\pi}{a}f(0). \eqno(B.19)
$$
The sense of going over from function ${G}_{\pm}(u)$ to functions $\tilde{G}_{\pm}(u)$ and $D_{\pm}(u)$ is 
that $\tilde{G}_{\pm}(-{\rm i}\kappa)+D_{\pm}(-{\rm i}\kappa)$, unlike ${G}_{\pm}(-{\rm i}\kappa)$, is exponentially 
decreasing at large values of $\kappa$. Defining 
$$
\Lambda(\kappa)=\frac{1}{4}[\tilde G_+(-{\rm i}\kappa)+D_+(-{\rm i}\kappa)+\tilde G_-(-{\rm i}\kappa)+D_-(-{\rm i}\kappa)], 
\eqno(B.20)
$$
which is explicitly given by (88), substituting the explicit form for $d_{+}(k)
+d_{-}(k)$ and recalling (B.5), we rewrite (B.19) into the form given by (87).
It should be noted that the contribution of poles on
the imaginary axis at $u=\pm{\rm i}(m\cos\varphi\sin\tilde\varphi + \cos\tilde\varphi\sqrt{2n|eB|+m^{2}\sin^{2}\varphi})$ 
and $u=\pm{\rm i}(m\cos\varphi\sin\tilde\varphi - \cos\tilde\varphi\sqrt{2n|eB|+m^{2}\sin^{2}\varphi})$ is cancelled.


\begin{thebibliography}{0}

\bibitem{Cho1}
A.Chodos, R.L.Jaffe, K.Johnson, C.B.Thorn and V.Weisskopf, {\it
Phys. Rev. D} {\bf 9}, 3471 (1974).

\bibitem{Cho2}
A.Chodos, R.L.Jaffe, K.Johnson and C.B.Thorn, {\it Phys. Rev. D}
{\bf 10}, 2599 (1974).

\bibitem{De}
T.De Grand, R.L.Jaffe, K.Johnson and J.Kiskis, {\it Phys. Rev. D}
{\bf 12}, 2060 (1975).

\bibitem{Joh}
K.Johnson, {\it Acta  Phys. Pol. B} {\bf 6}, 865 (1975).

\bibitem{Hei1}
W.Heisenberg, {\it Z. Phys.} {\bf 90}, 209 (1934).

\bibitem{Eul}
H.Euler and B.Kockel, {\it Naturwissensch.} {\bf 23}, 246 (1935).

\bibitem{Hei2}
W.Heisenberg and H.Euler, {\it Z. Phys.} {\bf 98}, 714 (1936).

\bibitem{Wei}
V.S.Weisskopf, {\it Kong. Dans. Vid. Selsk. Mat-Fys. Medd.} {\bf 14} No.6 (1936).

\bibitem{Schw}
J.Schwinger, {\it Phys. Rev.} {\bf 82}, 662 (1951).

\bibitem{Dun}
G.V.Dunne, 'Heisenberg-Euler effective lagrangians: Basics and extensions'. In: {\it Ian Kogan Memorial Collection 'From Fields to Strings: Circumnavigating Theoretical Physics'.} Ed. by M.Shifman, A.Vainshtein and J.Wheater (World Scientific, Singapore, 2004) vol.1, pp. 445-522.

\bibitem{Wie}
M.H.Al-Hashimi and U.-J.Wiese, {\it Annals Phys.} {\bf 327}, 1
(2012).

\bibitem{Si1}
Yu.A.Sitenko, {\it Phys. Rev. D} {\bf 91}, 085012 (2015).

\bibitem{Neu}
J.von Neumann, {\it Mathematische Grundlagen der Quantummechanik}
(Springer, Berlin, 1932).

\bibitem{Akhi}
N.I.Akhiezer and I.M.Glazman, {\it Theory of Linear Operators in
Hilbert Space} (Pitman, Boston, 1981).

\bibitem{Cas1}  H.B.G.Casimir, {\it Proc. Kon. Ned. Akad. Wetenschap
B} {\bf 51}, 793 (1948),

\bibitem{Mil} K.A.Milton, {\it The Casimir Effect: Physical Manifestations of
Zero-Point Energy} (World Scientific, River Edge, 2001).

\bibitem{Bor} M.Bordag, G.L.Klimchitskaya, U.Mohideen and V.M.Mostepanenko,
{\it Advances in the Casimir Effect} (Oxford University Press, Oxford, 2009).

\bibitem{Mer}
S.Mereghetti, {\it Astron. Astrophys. Rev.} {\bf 15}, 225 (2008).

\bibitem{Gra} 
D.Grasso and H.R.Rubinstein, {\it Phys. Rep.} {\bf 348}, 163 (2001).

\bibitem{Akhie}
A.I.Akhiezer and V.B.Berestetskij, {\it Quantum Electrodynamics}
(Interscience, New York, 1965).


\end{thebibliography}
\end{document}